\documentclass[twocolumn, onecolappendix, numberedappendix]{openjournal}

\renewcommand{\baselinestretch}{1.15}

\usepackage{lipsum}
\usepackage{hyperref}
\hypersetup{
    unicode, 
    colorlinks=true,
    linkcolor=linkcolor,
    citecolor=linkcolor,
    filecolor=linkcolor,
    urlcolor=linkcolor,
}
\usepackage{amsmath}
\usepackage{xcolor}
\definecolor{linkcolor}{HTML}{1E6173}
\usepackage{tensind}
\tensordelimiter{?}
\DeclareGraphicsExtensions{.bmp,.png,.jpg,.pdf}
\usepackage{verbatim}
\usepackage[normalem]{ulem}
\usepackage{orcidlink}
\usepackage{soul}
\usepackage{booktabs}
\usepackage{enumitem}
\usepackage{float}

\definecolor{keygreen}{RGB}{0,130,0}

\newcommand{\hmpc}{h^{-1}{\rm Mpc}}

\usepackage{lineno}
\graphicspath{ {./figs/} }

\begin{document}
\title[Towards practical field-level inference for weak lensing]{Towards practical field-level inference for weak lensing}

\author{\vspace{-4em}Yuuki Omori,$^{\ast,1,2,3}$
Justine Zeghal,$^{\dagger,4,5,6}$
Chihway Chang,$^{1,2,3}$
François Lanusse,$^{7}$\\
Laurence Perreault-Levasseur$^{4,5,6,8,9}$}

\affiliation{\vspace{-0.7em}}
\affiliation{$^{1}$Department of Astronomy and Astrophysics, University of Chicago, Chicago, IL 60637, USA}
\affiliation{$^{2}$Kavli Institute for Cosmological Physics, University of Chicago, Chicago, IL 60637, USA}
\affiliation{$^{3}$NSF-Simons AI Institute for the Sky (SkAI), 172 E. Chestnut St., Chicago, IL 60611, USA}

\affiliation{$^{4}$Department of Physics, Université de Montr\'{e}al, Montr\'{e}al H2V 0B3, Canada}
\affiliation{$^{5}$Mila – Quebec Artificial Intelligence Institute, Montr\'{e}al H2S 3H1, Canada}
\affiliation{$^{6}$Ciela – Montreal Institute for Astrophysical Data Analysis and Machine Learning, Montr\'{e}al H2V 0B3, Canada}
\affiliation{$^{7}$Universit\'{e} Paris-Saclay, Université Paris Cit\'{e}, CEA, CNRS, AIM, 91191 Gif-sur-Yvette, France}
\affiliation{$^{8}$CCA -- Flatiron Institute, 162 5th Ave, New York, NY 10010, USA}
\affiliation{$^{9}$Trottier Space Institute, Montreal, Quebec, Canada}

\email[$^\ast$]{yomori@uchicago.edu}
\email[$^\dagger$]{justine.zeghal@umontreal.ca}

\begin{abstract}
Nonlinear structure growth generates higher-order correlations and morphological features in the cosmic density field that cannot be fully characterized by two-point statistics. Upcoming surveys will measure these features with greater precision, making it essential to develop methods capable of extracting as much cosmological information as possible from them. Field-level inference (FLI) is one such approach, in which cosmological parameters are constrained by comparing observed maps to forward-modeled maps, either directly or through learned summaries that retain map-level information. In this work, we compare FLI with power-spectrum-based inference using the same forward-modeling pipeline for generating weak lensing maps, with the goal of quantifying the gain from map-level analysis relative to two-point statistics. We perform this comparison with both implicit and explicit inference methods, using 8-million-parameter forward models based on Lagrangian perturbation theory and particle-mesh (PM) $N$-body evolution. The two FLI approaches yield closely consistent posteriors; this agreement, together with coverage tests confirming the calibration of the implicit analyses, gives us confidence in the recovered field-level constraints. Relative to the power-spectrum-based analyses, these results show significant gains in cosmological information, especially when small scales are included in the PM-based forward model. We then discuss the remaining challenges that must be addressed before PM-based explicit FLI can be applied to observational datasets.
\end{abstract}


\maketitle

\section{Introduction}\label{sec:intro}
The distribution of matter in the Universe can be inferred through the effects of weak gravitational lensing, in which matter between distant galaxies and the observer perturbs the paths of light and induces small changes in their observed shapes. While the distortion of any individual galaxy is small, the coherent pattern measured across many background galaxies provides a statistical measure of the projected matter distribution~\citep[see][for reviews]{bartelmann2001,kilbinger2015,mandelbaum2018}. 

Weak lensing is most sensitive to large-scale structure roughly halfway between the observer and the source galaxies. For current optical galaxy surveys, this places the peak sensitivity at $z\sim0.5$. At these redshifts, gravitational collapse has already driven the density field into the nonlinear regime, where the statistics of the field differ significantly from those of a Gaussian random field. As a result, traditional two-point statistics, such as power spectra and correlation functions, do not fully capture the available information.

This has motivated extensive work on alternative summary statistics designed to capture non-Gaussian information beyond standard two-point functions, with the goal of extracting more cosmological information from existing datasets. These approaches range from natural extensions of two-point statistics to higher-order $N$-point statistics \citep[e.g., three-point mass aperture statistics,][]{semboloni2011,fu2014,secco2022,gomes2025,sugiyama2025} to more nonlinear summary statistics that capture combinations of many $N$-point functions, including peak statistics \citep{liux2015,kacprzak2016,harnoisderaps2021,marques2024}, Minkowski functionals \citep{kratochvil2012,grewal2022,armijo2025,novaes2025}, cumulative distribution functions \citep{banerjee2021,anbajagane2023}, scattering transforms \citep{cheng2021,cheng2025,gatti2024b}, and wavelet phase harmonics \citep{regaldosaintblancard2023,gatti2024b}.

A more general approach is field-level inference (FLI), in which cosmological parameters are inferred directly from the observed weak lensing field, exploiting its full spatial information. In this framework, the quantity of interest is the marginal posterior for the cosmological parameters,
\begin{equation}
p(\vec{\theta}|\vec{d}) = \int d\vec{u}\, p(\vec{\theta},\vec{u}|\vec{d}),
\end{equation}
where $\vec{d}$ denotes the observed lensing field and $\vec{u}$ denotes the whitened latent field specifying the realization of the initial conditions.\footnote{Although the field $\vec{u}$ is scientifically valuable as a probe of the initial conditions of the Universe prior to nonlinear structure growth, galaxy weak lensing provides relatively weak constraints on this field because of line-of-sight projection degeneracies. We therefore largely treat this latent space as a nuisance component in this work.} In \emph{explicit} inference, this marginal posterior is obtained by sampling the joint posterior over cosmological parameters and whitened latent field,
\begin{equation}\label{eq:jointpost}
p(\vec{\theta}, \vec{u}|\vec{d})\propto 
p(\vec{d}|\vec{\theta}, \vec{u})\, p(\vec{u})\, p(\vec{\theta}).
\end{equation}
The likelihood $p(\vec{d}|\vec{\theta}, \vec{u})$, which accounts for observational noise and survey effects, compares the observed lensing field $\vec{d}$ to the noiseless forward-model prediction
\begin{equation}
    \vec{m}=\mathcal{M}(\vec{\theta},\vec{u}).
\end{equation}
The cosmological parameters $\vec{\theta}$ determine the initial power spectrum and the growth of structure. For each sampled pair $(\vec{\theta},\vec{u})$, the forward model generates the corresponding initial density field, evolves it either analytically or numerically, and projects the resulting late-time density field along the line of sight to produce a weak lensing map. 

The central challenge of explicit inference stems from the need to sample over the latent field $\vec{u}$, whose number of degrees of freedom is set by the number of voxels used to represent the initial three-dimensional density field. Increasing the resolution of the simulation  improves the physical fidelity of the forward model, but also raises the computational cost. Conversely, insufficient resolution can introduce numerical artifacts that propagate through the inference, loss of cosmological information encoded in the observed weak lensing field, and degraded exploration of the posterior tails.

Despite these challenges, explicit FLI remains among the most statistically complete and interpretable approaches to extracting cosmological information from weak lensing maps. Although it relies on computational tools common in modern machine learning, such as automatic differentiation, GPU acceleration, and gradient-based optimization or sampling, the forward model itself is not a black box. Instead, the weak lensing field is generated from a physical forward model that evolves the density field and projects it into lensing observables. Beyond parameter constraints, explicit FLI yields the joint posterior over cosmology and the latent field, providing reconstructed density and convergence maps with quantified uncertainties. These field-level products enable cross-correlations with external datasets, and joint analyses of multiple probes that share the same underlying density field.
 
To date, studies carrying out explicit FLI for galaxy weak lensing have mainly relied on simplified statistical models of the weak lensing field rather than full gravitational evolution. For example, \citet{Boruah2024a, Boruah2024b, zeghal2025} approximate the two-dimensional galaxy weak lensing convergence field as a lognormal random field. This provides a tractable setting for explicit FLI, since the latent field is two-dimensional and can be sampled efficiently while still capturing some non-Gaussian features of the lensing field. Within this framework, they find that the information gain over two-point statistics depends on the parameter space considered: the improvement is modest for standard $\Lambda$CDM parameters, while in extensions such as $w$CDM, the non-Gaussian information can help break parameter degeneracies and lead to substantially tighter constraints.

A closely related and important precursor to our work is the series of analyses by \citet{Porqueres2022, Porqueres2023}, which used the Bayesian Origin Reconstruction from Galaxies framework \citep[BORG,][]{jasche2010,jasche2013} to perform explicit FLI for galaxy weak lensing. In this framework, the three-dimensional matter field is evolved using Lagrangian perturbation theory (LPT) and then integrated along the line of sight to generate a weak lensing field. This provides a more physically motivated forward model than a two-dimensional lognormal field, while remaining computationally tractable. They find significant information gains when moving from two-point statistics to FLI, leading to a qualitatively different conclusion from the lognormal analyses. Our explicit analysis follows the same broad approach of jointly inferring cosmological parameters and a three-dimensional latent matter field, which is evolved and projected into convergence maps, but differs in the forward model, implementation, and validation tests.

\emph{Implicit} FLI, on the other hand, avoids explicit sampling of the high-dimensional posterior by using the forward model only to generate a large ensemble of simulations, from which the marginal posterior is learned directly.\footnote{Here we focus on the marginal posterior but one can also learn the joint high-dimensional posterior.} This class of methods \citep[see][for a review]{cranmer2020} is commonly referred to as simulation-based inference (SBI), or likelihood-free inference, because the marginal likelihood $p(\vec{d}|\vec{\theta})$ is not written down or evaluated in closed form.  

In practice, implicit inference is usually carried out by first compressing the full lensing field into a lower-dimensional summary vector,
\begin{equation}
    \vec{y} = f_\varphi(\vec{d}),
\end{equation}
where $f_\varphi$ represents a learned compression that is approximately sufficient for the parameters $\vec{\theta}$. One then generates synthetic data realizations for different values of $\vec{\theta}$, with the latent initial conditions sampled internally by the simulator, and compresses each realization in the same way. The resulting pairs $(\vec{\theta},\vec{y})$ are used to train a neural density estimator, either to approximate the posterior $p(\vec{\theta}|\vec{y})$ directly \citep[Neural Posterior Estimation, e.g.,][]{blum2010non,papamakarios2016}, to approximate the likelihood $p(\vec{y}|\vec{\theta})$ \citep[Neural Likelihood Estimation, e.g.,][]{wood2010statistical,papamakarios2019}, or to estimate the likelihood ratio $p(\vec{y}|\vec{\theta})/p(\vec{y})$ \citep[Neural Ratio Estimation, e.g.,][]{cranmer2015approximating,thomas2022likelihood}. In the latter two cases, posterior samples are obtained through an additional sampling procedure using the learned likelihood or likelihood ratio.

Implicit inference has become widely used in cosmology because it can use high-fidelity simulation-based models without requiring an explicit analytic likelihood. In weak lensing, SBI has been applied both at the field level and with higher-order summary statistics such as peak counts, the convergence PDF, and Minkowski functionals, extracting information beyond standard two-point statistics while remaining computationally tractable for modern survey data. Such analyses have already been applied to KiDS-1000, HSC-Y1, and DES\,Y3 \citep{vonwietersheimkramsta2024,novaes2025,gatti2024a}.

While these studies have shown promising results, two caveats must be kept in mind:
\begin{itemize}[leftmargin=0.7em]
\item[-] Most practical SBI analyses require a compression step from the full data vector to a lower-dimensional representation $\vec{y}$. Depending on the chosen summary statistics or learned compression, this mapping may not be strictly information preserving, and some constraining power may be lost~\citep{akhmetzhanova2024,lanzieri2025}.
\item[-] When neural networks or neural density estimators are used either to compress the data or to approximate the posterior, the resulting inference must be carefully validated: posterior approximations can be miscalibrated or overconfident, and neural SBI methods can be sensitive to model misspecification~\citep{hermans2021,cannon2022}. In addition, their black-box nature can also make it difficult to interpret the learned mapping and to diagnose potential failure modes.
\end{itemize}

An important question is whether implicit and explicit FLI recover consistent cosmological constraints when applied to matched simulations and data. \citet{zeghal2025} showed that for 2D lognormal simulations and matched settings, the implicit and explicit approaches return the same cosmological constraints. Likewise, \citet{cuestalazaro2024}, using an alternative implicit inference that jointly recovers initial conditions and cosmological parameters, found that the posterior distributions of cosmological parameters obtained from explicit and implicit approaches agree closely when using dark-matter-only simulations.

In this work, we set up a controlled experiment to compare cosmological constraints from three approaches placed on equal footing: power spectrum inference, implicit FLI, and explicit FLI. In Section~\ref{sec:results}, we address the following questions:
\begin{enumerate}
\item {\it Can the posterior be correctly sampled using the explicit approach for LPT, and does this extend to the more realistic particle-mesh (PM) model? (Section \ref{sec:q1})}
\item {\it How much information is there in the lensing fields beyond two-point functions? (Section \ref{sec:q2}) }
\end{enumerate}

This paper is organized as follows. In Section~\ref{sec:weaklensing}, we provide a brief introduction to weak lensing and the background theory needed for this work. In Section~\ref{sec:methods}, we describe our implementation of the LPT- and PM-based forward models. In Section~\ref{sec:inference}, we describe the explicit and implicit FLI approaches considered in this work. We also describe the power-spectrum SBI analysis against which we compare the field-level results.  In Section~\ref{sec:results}, we present our main results, comparing the different modeling and inference approaches. Finally, we summarize and discuss the implications of this work and future improvements in Section~\ref{sec:discussion}.

Throughout this paper, we assume a fiducial cosmology with $\Omega_{\rm c}=0.266$, $\Omega_{\rm b}=0.0492$, $\sigma_{8}=0.831$, $h=0.673$, $n_{\rm s}=0.965$, and $w_{0}=-1$. In the cosmological inference, we vary $\Omega_{\rm m}$ and $S_{8}^{\rm lin}\equiv\sigma_{8}(\Omega_{\rm m}/0.3)^{\alpha}$, setting $\alpha=0.85$ for LPT and $\alpha=0.70$ for PM forward models, respectively.\footnote{The value of $\alpha$ is chosen to approximately follow the dominant degeneracy direction for each forward model. The appropriate scaling differs between LPT and PM because nonlinear structure growth introduces additional dependence on $\Omega_{\rm m}$ and $\sigma_8$ beyond the linear-amplitude scaling.}

\section{Galaxy weak lensing}\label{sec:weaklensing}

The weak lensing convergence field $\kappa$ is related to the two-dimensional lensing potential $\psi$ by
\begin{equation}
    \kappa = \frac{1}{2}\nabla^2 \psi,
\end{equation}
where we use the standard galaxy weak lensing convention.
Through the Poisson equation, we can rewrite the convergence field as a weighted line-of-sight projection of the density field (assuming a flat Universe):
\begin{equation}
\kappa (\hat{n}, \chi_s) = \frac{3 H_0^2 \Omega_{\rm m}}{2 c^2} \int_0^{\chi_{\rm s}} d\chi \frac{\chi (\chi_{\rm s} - \chi)}{a \chi_{\rm s}} \delta(\hat{n}, \chi),   
\label{eq:kappa}
\end{equation}
where $H_0$ and $\Omega_{\rm m}$ are the Hubble constant and the matter density parameter today, respectively. $c$ is the speed of light, $\chi_{\rm s}$ is the comoving distance to the source plane, $\delta$ is the matter overdensity, and $a$ is the cosmological scale factor. The convergence map that we observe is the integrated quantity:
\begin{equation}\label{eq:kappa_int}
\kappa(\hat{n})=\int_{0}^{\infty} d\chi_{\rm s}\, n(\chi_{\rm s})\, \kappa (\hat{n},\chi_{\rm s}),
\end{equation}
where $n(\chi)$ is the redshift distribution of the source galaxies.

\begin{table}
\begin{center}
\caption{Table summarizing the survey and simulation configurations, as well as the assumed cosmology. For the two sampled parameters, $\Omega_{\rm m}$ and $S_8^{\rm lin}$, we adopt uniform priors; the values in parentheses indicate the minimum and maximum prior bounds.}
\begin{tabular}{*4l} \toprule
{\bf Settings} & {\bf Value} \\ \midrule
\textbf{\textit{Survey configuration}} &  \\
%Area & 10$\times$10\,deg$^2$ \\
Num redshift bins & 2 \\
$\sigma_{\epsilon}$ & 0.26 \\
$n_{\rm gal}$ (per bin) & 10 arcmin$^{-2}$ \\
Mean redshift & [0.7, 0.8] \\
\midrule
\textbf{\textit{Simulation configuration}} &  \\
Box dimension & $450\times450\times2560$  $\hmpc$\\ 
Field dimension  & $10^{\circ}\times 10^{\circ}$ \\ 
Field grid & $144\times144\times384$ \\
Scale cuts ($\ell_{\rm max}$) & $400, 1000$ \\
\midrule
\textbf{\textit{Cosmological parameters}} \\
$\Omega_{\rm m}$ & $\mathcal{U}(0.1,1.0)$\\
$S_8^{\rm lin} \equiv \sigma_8 (\Omega_{\rm m}/0.3)^{\alpha}$  & $\mathcal{U}(0.4,1.4)$\\
$\alpha$ & 0.85 (LPT) or 0.70 (PM) \\
$h$ & 0.673 \\
$\Omega_{\rm b}$ & 0.0492 \\
$n_{s}$ & 0.965 \\
$w_{0}$ & -1 \\
\bottomrule
\hline
\end{tabular}

\label{tab:simsettings}
\end{center}
\end{table}

\section{Forward model}\label{sec:methods}

\subsection{Initial conditions}
We start by drawing cosmological parameters $\vec{\theta}=\{\Omega_{\rm m}, S_{8}^{\rm lin}\}$ from the priors listed in Table~\ref{tab:simsettings}. For each draw of $\vec{\theta}$, we generate a Gaussian realization of the linear density field by scaling the whitened latent field $\vec{u}$ by the square root of the linear matter power spectrum. In Fourier space, the coefficients of this field are denoted by $\widetilde{u}(\vec{k})$. The linear density field is then given by
\begin{equation}
\widetilde{s}(\vec{k}) = \sqrt{P(k,\vec{\theta})}\,\, \widetilde{u}(\vec{k}),
\end{equation}
where $P(k,\vec{\theta})$ is the linear matter power spectrum for parameters $\vec{\theta}$ at $z=0$. The field $\widetilde{s}(\vec{k})$ is therefore the present-day linear density field, with its amplitude at any earlier epoch set by the linear growth factor $D(a)$. For modes with a distinct conjugate partner,\footnote{Because the real-space field $u(\vec{x})$ is real, the Fourier coefficients obey Hermitian symmetry, $\widetilde{u}(-\vec{k})=\widetilde{u}^{*}(\vec{k})$. We therefore sample only the independent Fourier modes explicitly and set the conjugate modes by this symmetry. The DC mode is fixed to zero to enforce a vanishing mean overdensity.}
 we draw
\begin{equation}
\widetilde{u}(\vec{k}) = \frac{1}{\sqrt{2}}\left[a(\vec{k}) + i\, b(\vec{k})\right],
\qquad a,b \sim \mathcal{N}(0,1),
\end{equation}
so that $\langle |\widetilde{u}(\vec{k})|^2\rangle=1$. The real-space field is then obtained by an inverse Fourier transform,
\begin{equation}
s(\vec{x}) = \mathcal{F}^{-1}\!\left[\widetilde{s}(\vec{k})\right].
\end{equation}

From this linear field we set up the particle initial conditions in two stages. We first seed particles at $z=99$ ($a=0.01$), displacing them slightly from a uniform grid. We then advance the particles from $z=99$ to $z=15$ with a single \texttt{BullFrog} step \citep{rampf2025}, and the resulting $z=15$ field serves as the input to the subsequent gravitational evolution.

\subsection{Density evolution}

We implement both LPT and PM simulators in JAX~\citep{jax}, enabling automatic differentiation and efficient GPU execution. Differentiability provides exact gradients of the simulation output with respect to both the cosmological parameters and the $\mathcal{O}(10^7)$ latent-field modes, which makes gradient-based samplers such as Hamiltonian Monte Carlo \citep[HMC;][]{neal2011} and the No-U-Turn Sampler \citep[NUTS;][]{hoffman2014} applicable.

\subsubsection{LPT}
In 1LPT (or the Zel'dovich approximation), the final position of each particle is given by its initial (Lagrangian) position plus a displacement that grows with time according to linear theory. This can be written as:
\begin{equation}
\vec{x}(\vec{q}_{0},t)=\vec{q}_{0}+D(t)\Psi(\vec{q}_{0}),
\end{equation}
where $\vec{x}$ is the final position, $\vec{q}_0$ is the initial position, $D(t)$ is the growth factor, and $\Psi(\vec{q}_0)$ is the displacement field, which depends only on the initial density field. Since this is a simple analytic calculation, the displacement can be evaluated independently for each slice along the line of sight, using the growth factor $D$
 at that slice's redshift, to construct the lightcone.

While LPT has the benefit of being computationally light, it is accurate only in the linear regime. Once shell-crossing occurs, particle trajectories intersect and the approximation cannot correctly describe virialized halos. As a result, it underestimates small-scale power, fails to capture halo structure, and produces washed-out convergence maps. Modeling these low-redshift and small-scale regimes accurately therefore requires higher-order LPT, PM, or full $N$-body methods.

\subsubsection{PM: \texttt{BullFrog} integrator}
We evolve the matter field with a PM $N$-body simulator.\footnote{Our \texttt{BullFrog} implementation is closely informed by the implementation in \texttt{DISCO-DJ} \citep{list2025}, while the overall PM solver is closer in spirit to \texttt{JaxPM} combined with \texttt{BullFrog} time integration. This structure allows us to use anisotropic-resolution grids, which are needed to construct the lensing lightcone efficiently.} Particle positions and velocities are evolved with the \texttt{BullFrog} integrator \citep{list2024,rampf2025}, an LPT-informed time integrator built on a leapfrog structure that incorporates second-order LPT to achieve accurate evolution with a small number of time steps.

Particle evolution alternates between \textit{kick} steps, which update particle velocities using the gravitational forces, and \textit{drift} steps, which update particle positions using the current velocities. The main computational cost comes from the kick steps, because each kick requires a full particle-mesh force evaluation: the particle distribution is deposited onto a mesh, the Poisson equation is solved in Fourier space, and the resulting force field is interpolated back to the particle positions. Since this force solve must be repeated at every kick step, the runtime scales with the number of time steps, which motivates integrators such as \texttt{BullFrog} that retain accuracy with relatively few steps.

When computing the force, we deposit particles onto the mesh using a Triangular-Shaped Cloud (TSC) assignment kernel, and use the same kernel to interpolate fields back to the particle positions. The gravitational force is computed in Fourier space on a mesh whose resolution is increased by an integer factor $f_{\rm mesh}$ relative to the particle grid (we use $f_{\rm mesh}=3$). After TSC deposition, we deconvolve the corresponding assignment window, solve the Poisson equation, and compute each Cartesian component of the force by applying the Fourier-space gradient operator,
\begin{equation}
F_i(\boldsymbol{k}) = -{\rm i} k_i \Phi(\boldsymbol{k}).
\end{equation}
The resulting force field is transformed back to real space and interpolated to the particle positions, where it determines the momentum update in the kick step of the time integrator.

\subsection{Lensing}
\begin{figure}
\centering
\includegraphics[width=\linewidth]{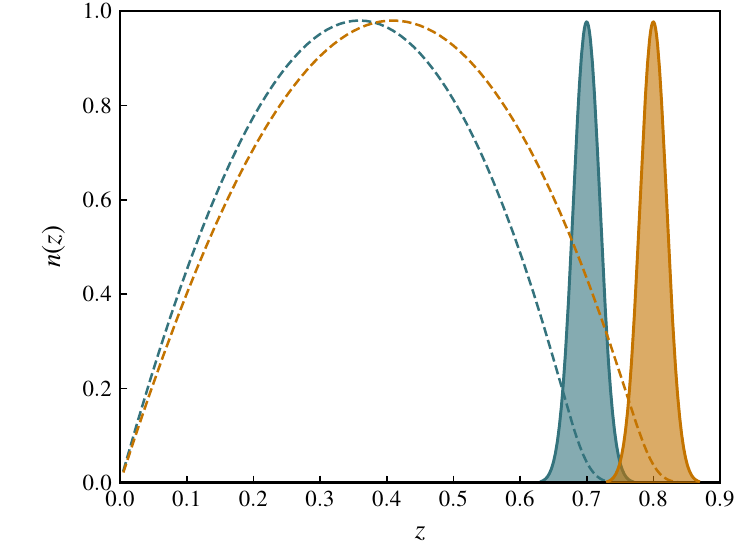}
\caption{Redshift distribution used in this work. The solid filled curves represent the source $n(z)$ at $z=0.7$ and $z=0.8$, while the dashed lines represent the lensing kernels.\\}
\label{fig:nofz}
\end{figure}

For this controlled study, we deliberately choose narrow source-redshift distributions centered at $z=0.7$ and $z=0.8$, as shown in Figure~\ref{fig:nofz}. This choice is motivated by the practical constraints of our forward models. At lower redshift, a fixed angular scale corresponds to smaller physical scales, which can fall below the simulation resolution. At higher redshift, the lightcone must cover a larger comoving volume, increasing the computational cost. The situation is further complicated by the cosmology dependence of the redshift-to-distance relation: across the prior range, the lensing kernel must remain fully contained within the simulated volume. The narrow redshift bins used here are therefore not intended to represent a realistic survey configuration, but rather to provide a controlled setup in which the forward model remains accurate, computationally feasible, and well defined throughout the parameter space explored in the inference.

\subsubsection{Born approximation}

We use the Born approximation to compute the weak lensing convergence from the density field of the LPT forward model. In practice, we adopt the discretized form of Equation~\ref{eq:kappa_int},
\begin{equation}\label{eq:kappa_slices}
\kappa(\hat{n}) = \sum_{i} W^{\kappa}(\chi_{i}) \delta(\hat{n},\chi_{i})\, \Delta\chi ,
\end{equation}
where
\begin{equation}
W^{\kappa}(\chi)=\frac{3H_{0}^{2}\Omega_{\rm m}}{2c^{2}}\frac{\chi}{a(\chi)}
\int_{\chi}^{\infty} d\chi_{\rm s}\, n(\chi_{\rm s}) \frac{\chi_{\rm s}-\chi}{\chi_{\rm s}},
\end{equation}
is the lensing kernel, and $\delta(\hat{n},\chi_{i})$ is the density contrast in the $i$-th radial shell.

\subsubsection{Ray tracing}

We use ray tracing to compute the weak lensing convergence from the density planes produced by the PM forward model. At each step, particles within thin comoving slices are projected onto two-dimensional grids using TSC mass assignment, producing projected density fields $\delta_i(\hat{n})$. Each plane is weighted by the lensing efficiency of Equation~\ref{eq:kappa} to give its convergence contribution $\kappa_i(\hat{n})$. The corresponding lensing potential $\psi_i$ is then obtained by solving the two-dimensional Poisson equation,
\begin{equation}
\nabla_\perp^2 \psi_{i} = 2\kappa_{i},
\end{equation}
which we solve in Fourier space. The deflection field on each plane is
\begin{equation}
\vec{\alpha}_{i} = \nabla_\perp \psi_{i},
\end{equation}
and these deflection fields are used to propagate rays through the sequence of lens planes, accounting for the change in each ray's trajectory as it crosses successive planes. The final convergence map is obtained from the distortion of the ray bundle.

We trace rays through the lens planes with a kick-drift-kick (KDK) leapfrog scheme that carries a two-component state: the angular position of the ray $\vec{\vartheta}$ and its angular velocity $\vec{\eta}$. At each plane the deflection bends the velocity (kick) and the ray then free-streams by its accumulated velocity (drift):
\begin{align}
\vec{\eta} &\;\leftarrow\; \vec{\eta} + \Delta\vec{\eta}^{\rm front}_i(\vec{\vartheta}), \\
\vec{\vartheta} &\;\leftarrow\; \vec{\vartheta} + f_{\rm drift}\,\vec{\eta}, \\
\vec{\eta} &\;\leftarrow\; \vec{\eta} + \Delta\vec{\eta}^{\rm back}_i(\vec{\vartheta}),
\end{align}
where $\Delta\vec{\eta}_i^{\,\mathrm{front}}$ and $\Delta\vec{\eta}_i^{\,\mathrm{back}}$ are the deflections from the front and back halves of slice $i$, each evaluated at the current ray position, and $f_{\mathrm{drift}}$ encodes the comoving-distance geometry across the slice. 

The ray-tracing procedure defines a lens mapping from the initial image-plane coordinate $\vec{\vartheta}_0$ to the final source-plane coordinate $\vec{\beta} \equiv \vec{\vartheta}_s$. The Jacobian of this mapping gives the main weak lensing observables used in this work,
\begin{align}
A_{ab} &= \frac{\partial \beta_a}{\partial \vartheta_{0,b}} =
\begin{pmatrix}
1-\kappa-\gamma_1 & -\gamma_2 \\
-\gamma_2 & 1-\kappa+\gamma_1
\end{pmatrix}.
\end{align}
Taking the trace of this relation gives the convergence,
\begin{align}
\kappa &= 1 - \tfrac{1}{2}\left(A_{11}+A_{22}\right).
\end{align}

\vspace{0.2cm}

\subsubsection{Noise}
\label{sec:noise}
In addition to the lensing signal, we include shape noise. We model the shape noise in each tomographic bin as an independent mean-zero Gaussian random field. For a pixelized map, the corresponding pixel-space variance is
\begin{equation}
\sigma_{\rm pix}^2
=
\frac{\sigma_\epsilon^2}{n_{\rm gal}\Omega_{\rm pix}},
\end{equation}
where \(\Omega_{\rm pix}\) is the pixel area, \(n_{\rm gal}\) is the source-galaxy number density, and \(\sigma_\epsilon\) is the intrinsic per-component ellipticity dispersion. We take $n_{\rm gal}=10\,{\rm gal}/{\rm arcmin}^{2}$ per tomographic bin and $\sigma_\epsilon=0.26$.

Since the Fourier transform is linear, this Gaussian noise model remains Gaussian when expressed in the two-dimensional Fourier plane. In the Fourier space likelihood used in Section \ref{sec:explicit}, the corresponding noise standard deviation is denoted by $\widetilde{\sigma}_{\rm shape}$, which is obtained from the same pixel-space variance after applying our discrete Fourier transform normalization. We assume a homogeneous survey depth and uncorrelated intrinsic galaxy shapes.

\subsubsection{Scale cuts}
Because our fast forward-modeling simulations are relatively low resolution, the grid and particle Nyquist scales are not far above the angular multipoles used for inference. We therefore take a conservative approach to ensure that finite-resolution effects, mass-assignment choices, and discretization errors from small-scale modes do not affect the analysis. To limit the impact of these numerical effects, we impose conservative $\ell$-space cuts and restrict the analysis to scales that are well resolved by the forward model.

We consider two choices, $\ell_{\rm max}=400$ and $\ell_{\rm max}=1000$. The conservative choice, $\ell_{\rm max}=400$, is designed to remove most of the nonlinear structure from the maps.\footnote{The $\ell_{\rm max}=400$ maps are not perfectly linear. Some nonlinear information remains at these scales, and field-level analyses can therefore still be mildly sensitive to information beyond the power spectrum.} The more aggressive choice, $\ell_{\rm max}=1000$, retains additional quasi-nonlinear information and allows us to test whether extending to smaller scales increases the information gain relative to the power-spectrum analysis, while still avoiding deeply nonlinear scales where simulation artifacts are more likely to affect the inference. This choice of $\ell_{\rm max}$ is broadly comparable to the maximum multipoles adopted in recent harmonic-space weak lensing analyses: the DES\,Y3 harmonic-space analysis uses tomographic-bin-dependent scale cuts with $\ell_{\rm max}\sim300$--$900$ \citep{doux2022}, while HSC-Y3 uses $\ell_{\rm max}=1800$ \citep{dalal2023} and the KiDS-Legacy band-power analysis uses $\ell_{\rm max}=1500$ \citep{wright2025}.  The nomenclature used in this work is summarized in Table \ref{tab:simalias}.

\begin{table}
\begin{center}
\caption{Summary of the simulation configurations and shorthand nomenclature.}
\begin{tabular}{ccccc} \toprule
Tag & Model & Inference  &  $\ell_{\rm max}$  & Data vector  \\\midrule
\texttt{LI4ff} & LPT & Implicit & 400 & Full field   \\
\texttt{LE4ff} & LPT & Explicit  & 400 & Full field   \\
\texttt{LI4ps} & LPT &  Implicit  &  400 & Power spectra \\
\texttt{LI10ff} & LPT & Implicit  & 1000 & Full field  \\
\texttt{LE10ff} & LPT & Explicit  & 1000 & Full field \\
\texttt{LI10ps} & LPT &  Implicit  &  1000 & Power spectra \\
\midrule
\texttt{PI4ff} & PM & Implicit & 400 & Full field  \\
\texttt{PE4ff} & PM & Explicit  & 400 & Full field  \\
\texttt{PI4ps} & PM & Implicit  & 400 & Power spectra \\
\texttt{PI10ff} & PM & Implicit & 1000 & Full field \\
\texttt{PE10ff} & PM & Explicit & 1000 & Full field \\
\texttt{PI10ps} & PM & Implicit & 1000 & Power spectra \\
\bottomrule
\end{tabular}
\label{tab:simalias}
\end{center}
\end{table}

\clearpage
\subsection{Validation and benchmark of forward model}
\label{sec:validation}
Table~\ref{tab:simrun} summarizes the computational cost of the two forward models on a single NVIDIA A100 GPU with 40\,GB of memory. Our implementation is optimized both numerically and through checkpointing to remain within this memory budget. This constraint is important in practice because 40\,GB A100s are more widely available than 80\,GB A100s or newer large-memory GPUs, making it feasible to run many independent chains in parallel.

\begin{figure}
\centering
\includegraphics[width=\linewidth]{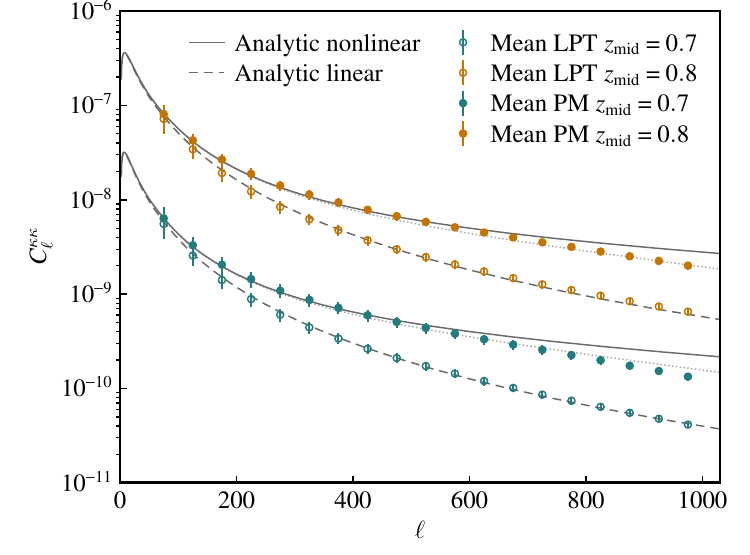}
\caption{Comparison of the mean recovered power spectra for the two source bins, plotted with the scatter from 2000 noiseless realizations. The solid and dashed lines show the analytical calculations using the same source-galaxy $n(z)$ and the nonlinear and linear matter power spectra, respectively. The dotted line shows an effective model in which the analytic nonlinear prediction is convolved with a $5'$ beam. The points and lines for the $z=0.8$ bin have been multiplied by a factor of 10 for visualization.\\ }
\label{fig:cl_comparison}
\end{figure}

In Figure~\ref{fig:cl_comparison}, we compare the noiseless power spectra measured from weak lensing maps generated from the LPT and PM density fields with two analytic predictions computed using \texttt{pyCCL}~\citep{chisari2019}. These predictions use the source-redshift distributions shown in Figure~\ref{fig:nofz} and are evaluated with either the linear or nonlinear matter power spectrum.

We find that the LPT simulations agree with the linear power-spectrum prediction on all scales considered, up to $\ell=1000$. The PM simulations are consistent with the nonlinear power spectrum up to $\ell\simeq 500$, but become biased low on smaller scales because of the finite resolution of the simulations. Matching the nonlinear analytic prediction on these scales would require either substantially higher-resolution simulations (which increases the number of forward model parameters and therefore the computational cost), or an on-the-fly small-scale correction procedure.\footnote{Several approaches for constructing ``upres'' simulations have been proposed in the literature, including neural-network-based corrections \citep{lanzieri2023} and optimal-transport-based methods \citep{zehgal2025b}.} We leave the implementation of these improvements to future work.

For the FLI presented here, the finite resolution of the forward model does not introduce the same kind of model mismatch that would arise when comparing to an external theory prediction, because the data realization is generated with the same forward model used in the likelihood. In this controlled setup, the finite-resolution response is therefore part of the generative model being tested. However, this response may still be cosmology dependent, and small-scale modes near the resolution limit can affect the inferred constraints if they are not adequately controlled. This also motivates the scale cuts described above.

\begin{figure}
\centering
\includegraphics[width=1.00\linewidth]{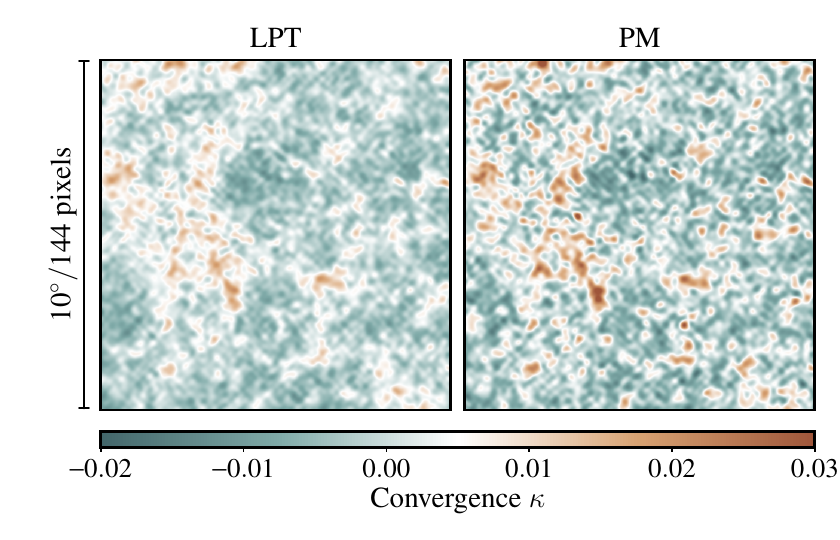}
\caption{Comparison of the noiseless convergence maps (second tomographic bin) between LPT and PM simulations prior to adding noise and applying the $\ell_{\rm max}$ cut.\\}
\label{fig:maps}
\end{figure}

\begin{table}
\begin{center}
\caption{Benchmarks of the simulation framework on an NVIDIA A100 with 40\,GB of memory. Numbers are averaged over 10 iterations. $N_{\rm steps}$/trajectory is the typical number of steps in the inference before hitting a U-turn.}
\begin{tabular}{ccc} \toprule
 & LPT & PM \\
\midrule
$N_{\rm part}$ & $384\times144\times144$ & $384\times144\times144$ \\
forward & 0.10s & 0.92s \\
gradient & 0.37s & 3.23s \\
leapfrog (1 step) & 0.75s & 6.45s \\ 
$N_{\rm steps}$/trajectory & 127 & 255 \\ 
gradient/forward ratio & 3.74 & 3.52 \\
peak mem & 8\,GB &  27\,GB\\
\bottomrule
\end{tabular}

\label{tab:simrun}
\end{center}
\end{table}

\section{Inference}\label{sec:inference}

\subsection{Power-spectrum-based implicit inference}
We first perform implicit inference based on the measured power spectra for each forward model, providing a two-point reference analysis for comparison with the corresponding field-level results. This allows us to isolate the information content of the two-point function while using the same SBI framework adopted for the implicit FLI analysis.

For each simulated map, we compute the binned auto- and cross-power spectra between the tomographic bins. These bandpowers define a summary vector,
\begin{equation}
\vec{t}_{\rm PS} = f_{\rm PS}(\vec{d}_{\rm sim}),
\end{equation}
where $\vec{d}_{\rm sim}$ denotes a noise-added map, $f_{\rm PS}$ denotes the operation of Fourier transforming the noisy lensing maps, applying the $\ell$-space scale cuts, and averaging the resulting auto- and cross-spectra in multipole bins. The final summary vector contains 21 bandpowers for $\ell_{\rm max}=400$ and 57 bandpowers for $\ell_{\rm max}=1000$. The same procedure is applied to both the training simulations and the data realization.

We then train a neural density estimator to approximate the posterior distribution $p(\vec{\theta}|\vec{t}_{\rm PS})$ using simulated pairs $(\vec{\theta},\vec{t}_{\rm PS})$. We use a conditional RealNVP flow \citep{realnvp} with 6 coupling layers and two-layer width-128 MLPs, trained on the binned auto- and cross-bandpowers from 40,000 simulations for 50,000 optimization steps.\footnote{For \texttt{PI4ps}, we use 5 coupling layers because the 6-layer model showed signs of overfitting.} Posterior samples are then drawn from the trained density estimator conditioned on the observed power-spectrum summary $\vec{t}_{\rm PS}^{\rm obs}=f_{\rm PS}(\vec{d})$.\\

\subsection{Explicit full-field inference}\label{sec:explicit}
We evaluate the likelihood $p(\vec{d}|\vec{\theta}, \vec{u})$ in Fourier space by comparing the real and imaginary parts of the observed and modeled convergence fields. Conditioned on the cosmological parameters and the whitened latent field, we assume that the Fourier-space residuals are Gaussian, so that
\begin{align}
-2&\ln\mathcal{L} \propto\nonumber\\ 
&\sum_{\vec{\ell}} \left[ \frac{ \left(\mathrm{Re}\left[\tilde{d}(\vec{\ell})-\tilde{m}(\vec{\ell})\right]\right)^2 }{\tilde{\sigma}_{\rm mode}^2(\vec{\ell})} + \frac{ \left(\mathrm{Im}\left[\tilde{d}(\vec{\ell})-\tilde{m}(\vec{\ell})\right]\right)^2 }{\tilde{\sigma}_{\rm mode}^2(\vec{\ell})} \right],
\end{align}
where the sum runs over the independent Fourier modes in the half-plane, and $\tilde{d}(\vec{\ell})$ and $\tilde{m}(\vec{\ell})$ are the Fourier transforms of the data and forward-modeled maps. The normalization $\tilde{\sigma}_{\rm mode}^2(\vec{\ell})$ is computed from the Fourier-space shape-noise variance. For generic complex modes, $\tilde{\sigma}_{\rm mode}^2(\vec{\ell}) = \tilde{\sigma}_{\rm shape}^2(\vec{\ell})/2$, while for purely real self-conjugate modes, such as the DC mode, $\tilde{\sigma}_{\rm mode}^2(\vec{\ell}) = \tilde{\sigma}_{\rm shape}^2(\vec{\ell})$.

In this construction, all non-Gaussianity and mode coupling in the signal are encoded in the forward model \(\widetilde m(\vec{\ell})\), while the likelihood only assumes that the remaining noise per mode is Gaussian and uncorrelated. Working in Fourier space also allows us to impose a sharp scale cut in \(\ell\)-space and avoids constructing and inverting a dense real-space covariance matrix, which would be prohibitively expensive.

We sample the joint posterior of Equation~\ref{eq:jointpost} using NUTS. The dominant obstacle is the strong posterior coupling between the cosmological parameters and a small set of coherent directions in the whitened latent space, which confines the posterior to a narrow ridge in the joint parameter space. We mitigate this with a rotation of the latent coordinates and a block-structured inverse mass matrix, whose construction is described in Appendix~\ref{app:precond}.

\subsection{Implicit full-field inference}
Our implicit FLI pipeline replaces direct evaluation of the field-level posterior with a neural density estimator that learns $p(\vec{\theta}|\vec{d})$ from simulated pairs $(\vec{\theta}, \vec{d}_{\rm sim})$. For both LPT and PM, the training pairs are generated with the same simulator used for explicit FLI. The pipeline has two stages. First, a ResNet18 compressor \citep{resnet} maps the tomographic convergence maps $\vec{d}_{\rm sim}$ of dimension $144 \times 144 \times 2$ to a 32-dimensional summary,
\begin{equation}
\vec{y} = f_\varphi(\vec{d}_{\rm sim}),
\qquad
\vec{y}\in{\mathbb R}^{32}.
\end{equation}
We use a summary dimension larger than the number of inferred parameters because empirical tests show that choices with $\text{dim}(\vec{y})\gg\text{dim}(\vec{\theta})$ yield tighter constraints than summaries whose dimension matches the number of inferred parameters.

The compressor is trained to produce summaries that retain as much information about $\vec{\theta}$ as possible. In the ideal limit, a sufficient summary would satisfy $I(\vec{y},\vec{\theta}) = I(\vec{d}_{\rm sim},\vec{\theta})$. In practice, we optimize a variational lower bound on this mutual information using the Variational Mutual Information Maximization loss function \citep[VMIM,][]{jeffrey2021}. Using the identity
\begin{equation}
I(\vec{y},\vec\theta)=
\mathbb{E}_{p(\vec{y},\vec\theta)}
\left[\log p(\vec\theta|\vec{y})\right]
+ H(\vec\theta),
\end{equation}
maximizing $I(\vec{y},\vec\theta)$ over $f_\varphi$ reduces to maximizing $\mathbb{E}[\log p(\vec\theta|\vec{y})]$. Because the exact conditional density is unavailable, VMIM introduces a variational density \(q_\psi(\vec\theta|\vec{y})\), giving the lower bound \citep{barber2003}
\begin{equation}
I(\vec{y},\vec\theta)
\geq
\mathbb{E}_{p(\vec{y},\vec\theta)}
\left[
\log q_\psi(\vec\theta|\vec{y})
\right]
+ H(\vec\theta).
\label{eq:variational_lower_bound}
\end{equation}
The compressor and variational density are trained jointly by minimizing
\begin{equation}
\mathcal{L}_{\rm VMIM}
=
-\mathbb{E}_{p(\vec{d}_{\rm sim},\vec\theta)}
\left[
\log q_\psi\!\left(\vec\theta|f_\varphi(\vec{d}_{\rm sim})\right)
\right],
\label{Eq:Loss_vmim}
\end{equation}
where the auxiliary variational density $q_{\psi}$ is a conditional RealNVP flow. The compressor is trained with 128,000 simulations, mini-batches of 128, and 400,000 optimization steps.

For each training batch, we generate $\vec{d}_{\rm sim}$ by injecting shape noise on the fly into the noiseless convergence maps using the same shape-noise model as in the explicit FLI analysis. This exposes the compressor and density estimator to multiple noise realizations for the same cosmological signal. For each batch, the maps are transformed with a 2D fast Fourier transform (FFT), independent Gaussian noise is added to the real and imaginary Fourier coefficients using the analytic survey noise level $\widetilde{\sigma}_{\rm shape}/\sqrt{2}$, and modes outside the analysis band, $\ell>\ell_{\max}$, are zeroed before transforming back to map space. We additionally apply random flips and rotations to the maps during compressor training.

After the compressor has been trained, its weights are frozen. We then train a separate neural posterior estimator on pairs \((\vec\theta,\vec{y})\), where \(\vec{y}=f_\varphi(\vec{d}_{\rm sim})\). The neural posterior \(q_\eta(\vec\theta\,|\,\vec{y})\) is learned by minimizing the negative log posterior
\begin{equation}
\mathcal{L}_{\rm NLL}
=
-\mathbb{E}_{p(\vec{y},\vec\theta)}
\left[
\log q_\eta(\vec\theta\,|\,\vec{y})
\right].
\label{eq:nll}
\end{equation}
The neural density estimator used is a conditional RealNVP flow, with four affine-coupling layers and coupling networks of width 128 and depth 4. It is trained for 80,000 optimization steps using 64,000 simulations. At inference time, the observed map is compressed, \(\vec{y}^{\rm obs}=f_\varphi(\vec{d})\), and posterior samples are drawn directly from
\begin{equation}
p(\vec\theta\,|\,\vec{d})
\simeq
q_\eta(\vec\theta\,|\,\vec{y}^{\rm obs}).
\end{equation}

\vspace{0.2cm}
\section{Results}\label{sec:results}

\begin{figure*}
    \centering
    \includegraphics[width=0.49\linewidth]{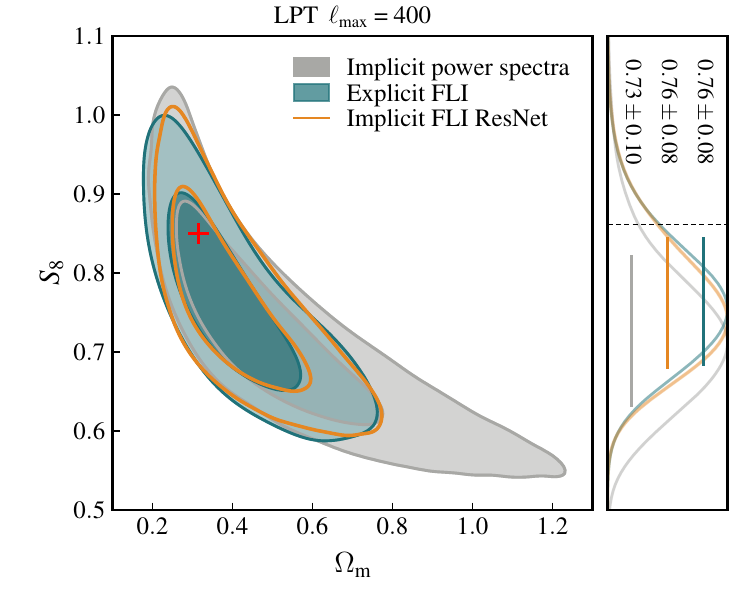}
    \includegraphics[width=0.49\linewidth]{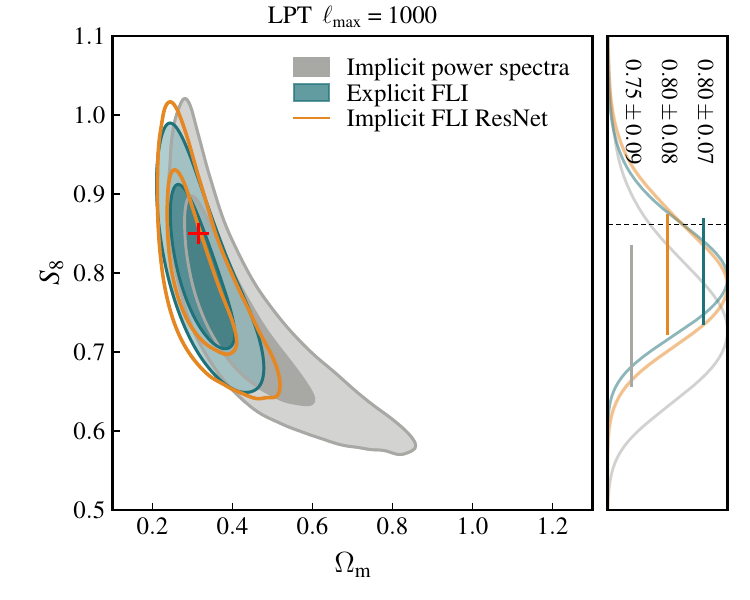}\\
    \includegraphics[width=0.49\linewidth]{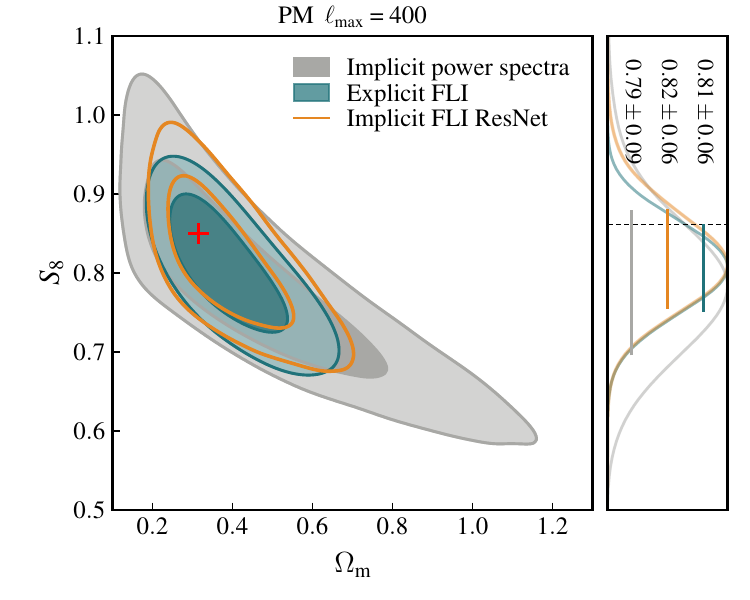}
    \includegraphics[width=0.49\linewidth]{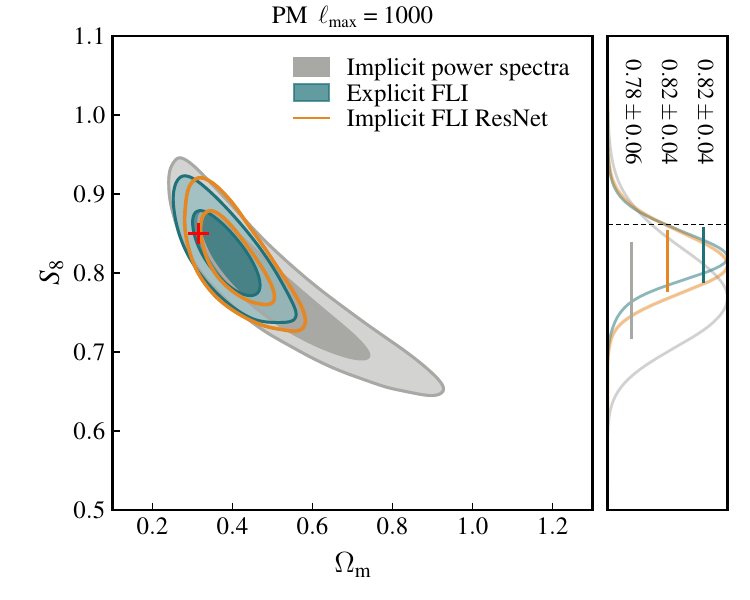}
    \caption{Cosmological constraints obtained using power-spectrum, implicit, and explicit inference with $\ell_{\rm max}=400$ and $\ell_{\rm max}=1000$ for LPT (upper panels) and PM (lower panels). Within each panel, all three approaches are applied to the same data realization. Note that the vertical axis is $S_{8}=\sigma_{8}(\Omega_{\rm m}/0.3)^{0.5}$, not $S_{8}^{\rm lin}=\sigma_{8}(\Omega_{\rm m}/0.3)^{\alpha}$, on which the inference is run.\\}
    \label{fig:results}
\end{figure*}

For each analysis, we report the marginal $1\sigma$ uncertainties on $\Omega_{\rm m}$ and $S_8^{\rm lin}$. However, because the degeneracy direction varies between setups, comparing the different analyses using only the one-dimensional marginalized uncertainties can be misleading. We therefore also report two measures of the two-dimensional constraining power. First, we use the Kullback--Leibler divergence between the posterior and the product of the flat priors on the two cosmological parameters,
\begin{align}
D_{\mathrm{KL}}\!\left(p \,\Vert\, \pi\right)
&= \int d\vec{\theta}\; p(\vec{\theta}\,|\,\vec{d})\,
\log_{2}\!\left[\frac{p(\vec{\theta}\,|\,\vec{d})}{\pi(\vec{\theta})}\right] \nonumber\\
&= \int d\vec{\theta}\; p(\vec{\theta}\,|\,\vec{d})\,\log_{2} p(\vec{\theta}\,|\,\vec{d}) + \log_{2} V \\
&\approx \frac{1}{N}\sum_{s=1}^{N} \log_{2} \hat{p}(\vec{\theta}_s) + \log_{2} V,
\end{align}
where we use $\pi(\vec{\theta}) = 1/V$ for the flat prior over volume $V$, and $\hat{p}(\vec{\theta}_s)$ is a kernel density estimate of the posterior evaluated at each posterior sample $\vec{\theta}_s$. We also report the figure of merit (FoM):
\begin{equation}
{\rm FoM} = \frac{1}{\sqrt{\det \Sigma_{\Omega_{\rm m}, S_{8}^{\rm lin}}}},
\end{equation}
where $\Sigma_{\Omega_{\rm m}, S_{8}^{\rm lin}}$ is the posterior covariance of $\Omega_{\rm m}$ and $S_8^{\rm lin}$. The FoM captures only the Gaussian, local constraint volume, whereas $D_{\rm KL}$ is sensitive to the full, possibly non-Gaussian, posterior shape.\\

\begin{table}
\begin{center}
\caption{Summary of our main results: uncertainty in $\Omega_{\rm m}$, $S_{8}^{\rm lin}$, FoM and $D_{\rm KL}$. We also quote the improvement relative to power spectrum analysis for the implicit case.}
\begin{tabular}{lrrrrr}
\hline
 Run             &   $\Delta \Omega_{\rm m}$ &   $\Delta S_{8}^{\rm lin}$ &   FoM &   $D_{\mathrm{KL}}$ [bits] &   \%vs PS \\
\hline
 \texttt{LI4ps}  &                     0.241 &                      0.054 &    81 &                       2.83 &         - \\
 \texttt{LI10ps} &                     0.158 &                      0.044 &   144 &                       3.80 &         - \\
 \texttt{PI4ps}  &                     0.225 &                      0.052 &    86 &                       2.93 &         - \\
 \texttt{PI10ps} &                     0.154 &                      0.030 &   246 &                       4.32 &         - \\
\hline
 \texttt{LI4ff}  &                     0.120 &                      0.057 &   149 &                       3.66 &     +77\% \\
 \texttt{LI10ff} &                     0.063 &                      0.051 &   310 &                       4.77 &     +96\% \\
 \texttt{PI4ff}  &                     0.107 &                      0.046 &   202 &                       4.04 &    +116\% \\
 \texttt{PI10ff} &                     0.061 &                      0.029 &   564 &                       5.47 &    +123\% \\
\hline
 \texttt{LE4ff}  &                     0.118 &                      0.051 &    -- &                         -- &        -- \\
 \texttt{LE10ff} &                     0.053 &                      0.044 &    -- &                         -- &        -- \\
 \texttt{PE4ff}  &                     0.098 &                      0.039 &    -- &                         -- &        -- \\
 \texttt{PE10ff} &                     0.059 &                      0.024 &    -- &                         -- &        -- \\
\hline
\end{tabular}
\label{tab:results}
\end{center}
\end{table}

\begin{table}
\begin{center}
\caption{Convergence diagnostics for the explicit field-level chains. For each parameter we report the bulk effective sample size, with the tail ESS (5th/95th percentile) in parentheses, and the rank-normalized $\hat{R}$ convergence diagnostic.}
\begin{tabular}{lrrrrr}
\hline
 Run             &   $n_{\rm samp}$ &   $\Omega_{\rm m}$ ESS &   $\Omega_{\rm m}$ $\hat{R}$ &   $S_{8}^{\rm lin}$ ESS &   $S_{8}^{\rm lin}$ $\hat{R}$ \\
\hline
 \texttt{LE4ff}  &         $10^{4}$ &               80 (108) &                         1.09 &             2183 (3195) &                          1.00 \\
 \texttt{LE10ff} &         $10^{4}$ &                42 (83) &                         1.16 &               596 (829) &                          1.01 \\
 \texttt{PE4ff}  &             6200 &               42 (168) &                         1.30 &             1681 (3628) &                          1.01 \\
 \texttt{PE10ff} &             5200 &                29 (51) &                         1.53 &               286 (914) &                          1.04 \\
\hline
\end{tabular}
\label{tab:explicit_convergence}
\end{center}
\end{table}

\subsection{Sampling in high-dimensional parameter space}\label{sec:q1}

We first assess the robustness of the implicit posteriors. We do this by performing Tests of Accuracy with Random Points \citep[TARP;][]{lemos2023}, described in Appendix~\ref{app:tarp} and shown in Figure~\ref{fig:tarp}. This test checks whether nominal credible regions contain the true parameters with the expected frequency across simulations. For a calibrated posterior, the empirical coverage should follow the diagonal relation between nominal credibility and expected coverage. The curves for the implicit field-level analyses are highly consistent with the diagonal within the realization-to-realization scatter, indicating that the NPE posteriors are not strongly miscalibrated. This supports their use as a reference for the explicit inference comparison, while also emphasizing that the implicit constraints should be interpreted as calibrated within the precision of the finite simulation suite rather than as an exact ground truth. We also test an alternative, simpler convolutional neural network (CNN) compressor in Appendix~\ref{app:cnn}, finding consistent degeneracy directions but marginally looser constraints, as expected for a less expressive architecture.

Next, we compare the results between implicit and explicit inference. For LPT, we compare \texttt{LE4ff}/\texttt{LE10ff} with \texttt{LI4ff}/\texttt{LI10ff}; for PM, we compare \texttt{PE4ff}/\texttt{PE10ff} with \texttt{PI4ff}/\texttt{PI10ff}. The 2D posteriors are broadly consistent in shape, recovering similar degeneracy directions in the $(\Omega_{\rm m}, S_8^{\rm lin})$ plane. Comparing the marginal $1\sigma$ uncertainties, the two approaches give similar constraints on both parameters. For the LPT runs, $\Delta\Omega_{\rm m} = 0.118$ and $0.053$ for the explicit analyses, compared to $0.120$ and $0.063$ for the implicit analyses, for $\ell_{\rm max}=400$ and $1000$, respectively. The corresponding uncertainties are $\Delta S_8^{\rm lin} = 0.051$ and $0.044$ for the explicit analyses, compared to $0.057$ and $0.051$ for the implicit analyses. The PM runs show a similar level of agreement: $\Delta\Omega_{\rm m} = 0.098$ and $0.059$ for the explicit analyses, compared to $0.107$ and $0.061$ for the implicit analyses, and $\Delta S_8^{\rm lin} = 0.039$ and $0.024$, compared to $0.046$ and $0.029$. In all cases, the explicit constraints are comparable to or slightly tighter than the implicit ones, with the improvement most reliable in the well-converged $S_8^{\rm lin}$ direction.

Finally, for the explicit inference runs, we report the rank-normalized $\hat{R}$ statistic and effective sample size (ESS), with the results summarized in Table~\ref{tab:explicit_convergence}. The $S_8^{\rm lin}$ direction is well sampled in all chains, with $\hat{R}$ close to unity, while the $\Omega_{\rm m}$ direction mixes more slowly. The LPT runs have $\hat{R}=1.09$ and $1.16$ for $\Omega_{\rm m}$, while the PM runs have $\hat{R}=1.30$ and $1.53$, with bulk ESS values of only a few tens. We therefore do not compare the FoM or $D_{\rm KL}$ between the explicit and implicit runs, since these quantities depend on the full width and orientation of the $\Omega_{\rm m}$--$S_8^{\rm lin}$ degeneracy. We instead use the explicit runs as a consistency check on the recovered degeneracy direction and on the well-converged $S_8^{\rm lin}$ uncertainty. Their agreement with the implicit posteriors indicates that the learned compression used in the implicit analysis does not discard a large amount of field-level information. The explicit PM run also demonstrates that the field-level posterior can be sampled directly, although achieving full convergence in the $\Omega_{\rm m}$ direction remains an important target for future work.\\

\subsection{Information beyond two-point functions}\label{sec:q2}
We compare the implicit field-level analyses to the corresponding power-spectrum analyses to quantify the information gained beyond a two-point-function analysis. As described in Section~\ref{sec:inference}, this comparison is performed using the same data realization, priors, cosmological parameter space, and forward simulations; only the compression of the maps differs. Since both sets of posteriors are obtained as independent draws from trained density estimators, their FoM and $D_{\rm KL}$ values can be compared directly.

Table~\ref{tab:results} summarizes the comparison. For the LPT model, the power-spectrum baselines give ${\rm FoM}=81$ and $144$ for \texttt{LI4ps} and \texttt{LI10ps}, with $D_{\rm KL}=2.83$ and $3.80$ bits. Implicit FLI reaches ${\rm FoM}=149$ and $310$, with $D_{\rm KL}=3.66$ and $4.77$ bits. The corresponding information-volume gains, defined as $2^{\Delta D_{\rm KL}}-1$, are  $77\%$ and $96\%$ for $\ell_{\rm max}=400$ and $1000$, respectively. The same pattern, with a larger gain, holds for the PM model. The power-spectrum baselines give ${\rm FoM}=86$ and $246$ for \texttt{PI4ps} and \texttt{PI10ps}, with $D_{\rm KL}=2.93$ and $4.32$ bits. The implicit FLI increases these to ${\rm FoM}=202$ and $564$, with $D_{\rm KL}=4.04$ and $5.47$ bits. These correspond to information-volume gains of $116\%$ and $123\%$. The field-level gain is larger for PM than for LPT, consistent with the PM forward model capturing nonlinear structure that the LPT model does not, and it grows as the analysis extends to smaller scales ($\ell_{\rm max}=1000$ versus $400$).

These results show that the field-level analyses extract substantially more information from the same maps than the power-spectrum analyses. Since the implicit pipeline first compresses the map into a learned summary, the gains measured here are a conservative estimate of the information available in the full field. The slightly tighter explicit constraints in the well-converged $S_8^{\rm lin}$ direction are consistent with this interpretation.\\

\section{Summary and outlook}\label{sec:discussion}

In this work, we compared power-spectrum inference, implicit FLI, and explicit FLI using the same weak lensing forward-modeling framework. We used this comparison to test whether implicit and explicit approaches give consistent cosmological posteriors, and to quantify how much additional cosmological information can be extracted when convergence maps are analyzed at the field level rather than through angular power spectra. To isolate differences between inference strategies and data compression choices, we used matched data realizations, priors, cosmological parameter space, and LPT- and PM-based forward models across the corresponding analyses. Our main conclusions are summarized below.

\begin{itemize}[leftmargin=1.2em,parsep=0.0em]

\item[-] The implicit and explicit field-level constraints are broadly consistent in the $(\Omega_{\rm m},S_{8}^{\rm lin})$ plane, and yield similar marginalized $1\sigma$ uncertainties at the one-dimensional level. This agreement, together with the TARP coverage tests showing that the implicit posteriors are well calibrated within the precision of the finite simulation suite, gives confidence that the recovered posteriors provide a reliable estimate of the field-level constraints attainable in this setup.

\item[-] Direct explicit sampling of the PM field-level posterior is feasible, but remains computationally challenging. The explicit runs recover the expected degeneracy direction and the well-converged $S_8^{\rm lin}$ uncertainty, with contours that are broadly consistent with the implicit PM analysis. However, the convergence diagnostics indicate persistent slow mixing in $\Omega_{\rm m}$, even after applying the coupling-direction rotation and block-structured inverse mass matrix.

\item[-] The block-structured inverse mass matrix is essential for making the explicit high-dimensional sampler practical. By identifying the low-dimensional subspace in which cosmological parameters couple coherently to the whitened latent field, the sampler can assign a dense inverse mass matrix only to the small $(\vec{\theta},\vec{a})$ block while keeping the remaining $\sim$$10^7$ latent-field directions diagonal. This substantially improves the geometry seen by NUTS without requiring a full dense mass matrix in the field-level parameter space.

\item[-] Comparing the implicit FLI results against the corresponding power-spectrum implicit analyses, whose posteriors are also independent draws from trained density estimators and are therefore not subject to sampling convergence limitations, we find improvements of $77\%$ and $96\%$ for the LPT forward model at $\ell_{\rm max}=400$ and $1000$, and $116\%$ and $123\%$ for the PM forward model. Because the implicit field-level analysis uses a compressed representation of the map, these gains should be viewed as a conservative estimate of the improvement available from field-level information.

\end{itemize}

Running explicit FLI on observational data will require PM-based forward models, since LPT fails to reproduce nonlinear structure even at $\ell=400$. While the issue of slow mixing in $\Omega_{\rm m}$ could in principle be overcome simply by running longer chains, in practice, reaching convergence within a reasonable time will require a better parameterization of the problem or more efficient sampling schemes. The PM-based explicit FLI framework described in this work is therefore a demonstration of feasibility rather than a pipeline ready for real-data analysis. Applying it to upcoming surveys will additionally require larger cosmological volumes, realistic source-redshift distributions with more tomographic bins, and higher forward-model resolution to reach the $\ell_{\rm max}$ values used in those analyses, all of which increase the cost of each forward and gradient evaluation and compound the sampling challenge described above.

However, meeting these requirements should be within reach as more capable hardware becomes widely available and software optimization continues. Higher-memory GPUs would reduce the memory pressure of each PM evaluation and lessen the need for memory-saving strategies such as checkpointing and recomputation of intermediate states, which trade runtime for memory. The single-GPU volume limitation could be overcome by distributing one simulation across multiple GPUs; \texttt{jaxDecomp} \citep{kabalan2026} was developed for this purpose, making larger-volume PM forward models feasible. Together with low-level optimizations of the PM evolution and ray-tracing operations, as well as further improvements to the inverse mass matrix or learned proposals, these developments should bring explicit FLI closer to practical use on observational datasets.

This study was restricted to an idealized weak-lensing setup in which shape noise was the only observational effect considered. A natural extension is to include additional physical and observational effects that must be modeled in real data, such as intrinsic alignments, baryonic effects, masking, photometric-redshift uncertainties, and survey systematics. Once these effects are included, explicit FLI should be relevant for weak lensing data from Rubin, \emph{Euclid}, and \emph{Roman}, in particular their deep-field observations, where high source density and small statistical errors allow clean measurement of the non-Gaussian features, from which substantial cosmological information can be extracted.

\section*{Acknowledgments}
We thank Alan Zhou and Benjamin Remy for helpful discussions. YO and CC were supported by NSF grant AST-2406551. We acknowledge the support from the France and Chicago Collaborating in the Sciences (FACCTS) award, which seeded the initial collaboration that led to this paper.

This research used resources of the Argonne Leadership Computing Facility, which is a U.S. Department of Energy Office of Science User Facility operated under contract DE-AC02-06CH11357. This research used resources of the National Energy Research Scientific Computing Center (NERSC), a Department of Energy User Facility using NERSC awards HEP-ERCAP 0038417 and HEP-ERCAP 0035553. We gratefully acknowledge the computing resources provided on Crossover, a high-performance computing cluster operated by the Laboratory Computing Resource Center at Argonne National Laboratory. We acknowledge the University of Chicago’s Research Computing Center for their support of this work. 

JZ and LPL were supported by the Simons Collaboration on ‘Learning the Universe’ and by Schmidt Sciences, a philanthropic initiative founded by Eric and Wendy Schmidt as part of the Virtual Institute for Astrophysics (VIA). This work is in part supported by computational resources provided by Calcul Quebec and the Digital Research Alliance of Canada. This work was granted access to the HPC/AI resources of IDRIS under the allocations AD011014029R1 made by GENCI.

\clearpage
\bibliographystyle{apsrev4-2-trim}
\bibliography{oja_template}

@ARTICLE{akhmetzhanova2024,
       author = {{Akhmetzhanova}, Aizhan and {Mishra-Sharma}, Siddharth and {Dvorkin}, Cora},
        title = "{Data compression and inference in cosmology with self-supervised machine learning}",
      journal = {\mnras},
         year = 2024,
        month = jan,
       volume = {527},
       number = {3},
        pages = {7459-7481},
          doi = {10.1093/mnras/stad3646},
archivePrefix = {arXiv},
       eprint = {2308.09751},
 primaryClass = {astro-ph.CO},
       adsurl = {https://ui.adsabs.harvard.edu/abs/2024MNRAS.527.7459A}
}

@article{realnvp,
  title={Density estimation using real nvp},
  author={Dinh, Laurent and Sohl-Dickstein, Jascha and Bengio, Samy},
  journal={arXiv preprint arXiv:1605.08803},
  year={2016}
}

@inproceedings{resnet,
  title={Deep residual learning for image recognition},
  author={He, Kaiming and Zhang, Xiangyu and Ren, Shaoqing and Sun, Jian},
  booktitle={Proceedings of the IEEE conference on computer vision and pattern recognition},
  pages={770--778},
  year={2016}
}

@ARTICLE{anbajagane2023,
       author = {{Anbajagane}, D. and {Chang}, C. and {Banerjee}, A. and others},
        title = "{Beyond the 3rd moment: a practical study of using lensing convergence CDFs for cosmology with DES Y3}",
      journal = {\mnras},
         year = 2023,
        month = dec,
       volume = {526},
       number = {4},
        pages = {5530-5554},
          doi = {10.1093/mnras/stad3118},
archivePrefix = {arXiv},
       eprint = {2308.03863},
 primaryClass = {astro-ph.CO},
       adsurl = {https://ui.adsabs.harvard.edu/abs/2023MNRAS.526.5530A}
}

@ARTICLE{armijo2025,
       author = {{Armijo}, Joaquin and {Marques}, Gabriela A. and {Novaes}, Camila P. and {Thiele}, Leander and {Cowell}, Jessica A. and {Grand{\'o}n}, Daniela and {Shirasaki}, Masato and {Liu}, Jia},
        title = "{Cosmological constraints using Minkowski functionals from the first year data of the Hyper Suprime-Cam}",
      journal = {\mnras},
         year = 2025,
        month = mar,
       volume = {537},
       number = {4},
        pages = {3553-3560},
          doi = {10.1093/mnras/staf257},
archivePrefix = {arXiv},
       eprint = {2410.00401},
 primaryClass = {astro-ph.CO},
       adsurl = {https://ui.adsabs.harvard.edu/abs/2025MNRAS.537.3553A}
}

@ARTICLE{banerjee2021,
       author = {{Banerjee}, Arka and {Abel}, Tom},
        title = "{Cosmological cross-correlations and nearest neighbour distributions}",
      journal = {\mnras},
         year = 2021,
        month = jun,
       volume = {504},
       number = {2},
        pages = {2911-2923},
          doi = {10.1093/mnras/stab961},
archivePrefix = {arXiv},
       eprint = {2102.01184},
 primaryClass = {astro-ph.CO},
       adsurl = {https://ui.adsabs.harvard.edu/abs/2021MNRAS.504.2911B}
}

@article{barber2003,
  title={Information maximization in noisy channels: A variational approach},
  author={Barber, David and Agakov, Felix},
  journal={Advances in Neural Information Processing Systems},
  volume={16},
  year={2003}
}

@ARTICLE{bartelmann2001,
       author = {{Bartelmann}, M. and {Schneider}, P.},
        title = "{Weak gravitational lensing}",
      journal = {\physrep},
         year = 2001,
        month = jan,
       volume = {340},
       number = {4-5},
        pages = {291-472},
          doi = {10.1016/S0370-1573(00)00082-X},
archivePrefix = {arXiv},
       eprint = {astro-ph/9912508},
 primaryClass = {astro-ph},
       adsurl = {https://ui.adsabs.harvard.edu/abs/2001PhR...340..291B}
}

@article{blum2010non,
  title={Non-linear regression models for Approximate Bayesian Computation},
  author={Blum, Michael GB and Fran{\c{c}}ois, Olivier},
  journal={Statistics and computing},
  volume={20},
  number={1},
  pages={63--73},
  year={2010},
  publisher={Springer}
}

@ARTICLE{boruah2024a,
        author = {{Boruah}, Supranta S. and {Rozo}, Eduardo},
         title = "{Map-based cosmology inference with weak lensing - information content and its dependence on the parameter space}",
       journal = {\mnras},
          year = 2024,
         month = jan,
        volume = {527},
        number = {1},
         pages = {L162-L166},
           doi = {10.1093/mnrasl/slad160},
archivePrefix = {arXiv},
        eprint = {2307.00070},
  primaryClass = {astro-ph.CO},
        adsurl = {https://ui.adsabs.harvard.edu/abs/2024MNRAS.527L.162B}
}

@ARTICLE{boruah2024b,
        author = {{Boruah}, Supranta S. and {Fiedorowicz}, Pier and {Rozo}, Eduardo},
         title = "{Bayesian mass mapping with weak lensing data using KARMMA -- validation with simulations and application to Dark Energy Survey Year 3 data}",
       journal = {arXiv e-prints},
          year = 2024,
         month = mar,
           eid = {arXiv:2403.05484},
         pages = {arXiv:2403.05484},
           doi = {10.48550/arXiv.2403.05484},
archivePrefix = {arXiv},
        eprint = {2403.05484},
  primaryClass = {astro-ph.CO},
        adsurl = {https://ui.adsabs.harvard.edu/abs/2024arXiv240305484B}
}

@software{jax,
  author = {James Bradbury and Roy Frostig and Peter Hawkins and others},
  title = {{JAX}: composable transformations of {P}ython+{N}um{P}y programs},
  url = {http://github.com/jax-ml/jax},
  version = {0.3.13},
  year = {2018},
}

@ARTICLE{cannon2022,
       author = {{Cannon}, Patrick and {Ward}, Daniel and {Schmon}, Sebastian M.},
        title = "{Investigating the Impact of Model Misspecification in Neural Simulation-based Inference}",
      journal = {arXiv e-prints},
         year = 2022,
        month = sep,
          eid = {arXiv:2209.01845},
        pages = {arXiv:2209.01845},
          doi = {10.48550/arXiv.2209.01845},
archivePrefix = {arXiv},
       eprint = {2209.01845},
 primaryClass = {stat.ML},
       adsurl = {https://ui.adsabs.harvard.edu/abs/2022arXiv220901845C}
}

@ARTICLE{cheng2021,
        author = {{Cheng}, Sihao and {M{\'e}nard}, Brice},
         title = "{Weak lensing scattering transform: dark energy and neutrino mass sensitivity}",
       journal = {\mnras},
          year = 2021,
         month = oct,
        volume = {507},
        number = {1},
         pages = {1012-1020},
           doi = {10.1093/mnras/stab2102},
archivePrefix = {arXiv},
        eprint = {2103.09247},
  primaryClass = {astro-ph.CO},
        adsurl = {https://ui.adsabs.harvard.edu/abs/2021MNRAS.507.1012C}
}

@ARTICLE{cheng2025,
       author = {{Cheng}, Sihao and {Marques}, Gabriela A. and {Grand{\'o}n}, Daniela and others},
        title = "{Cosmological constraints from weak lensing scattering transform using HSC Y1 data}",
      journal = {\jcap},
         year = 2025,
        month = jan,
       volume = {2025},
       number = {1},
          eid = {006},
        pages = {006},
          doi = {10.1088/1475-7516/2025/01/006},
archivePrefix = {arXiv},
       eprint = {2404.16085},
 primaryClass = {astro-ph.CO},
       adsurl = {https://ui.adsabs.harvard.edu/abs/2025JCAP...01..006C}
}

@ARTICLE{chisari2019,
        author = {{Chisari}, Nora Elisa and {Alonso}, David and {Krause}, Elisabeth and others},
         title = "{Core Cosmology Library: Precision Cosmological Predictions for LSST}",
       journal = {\apjs},
          year = 2019,
         month = may,
        volume = {242},
        number = {1},
           eid = {2},
         pages = {2},
           doi = {10.3847/1538-4365/ab1658},
archivePrefix = {arXiv},
        eprint = {1812.05995},
  primaryClass = {astro-ph.CO},
        adsurl = {https://ui.adsabs.harvard.edu/abs/2019ApJS..242....2C}
}

@article{cranmer2015approximating,
  title={Approximating likelihood ratios with calibrated discriminative classifiers},
  author={Cranmer, Kyle and Pavez, Juan and Louppe, Gilles},
  journal={arXiv preprint arXiv:1506.02169},
  year={2015}
}

@ARTICLE{cranmer2020,
       author = {{Cranmer}, Kyle and {Brehmer}, Johann and {Louppe}, Gilles},
        title = "{The frontier of simulation-based inference}",
      journal = {Proceedings of the National Academy of Science},
         year = 2020,
        month = dec,
       volume = {117},
       number = {48},
        pages = {30055-30062},
          doi = {10.1073/pnas.1912789117},
archivePrefix = {arXiv},
       eprint = {1911.01429},
 primaryClass = {stat.ML},
       adsurl = {https://ui.adsabs.harvard.edu/abs/2020PNAS..11730055C}
}

@inproceedings{cuestalazaro2024,
  title={Joint cosmological parameter inference and initial condition reconstruction with Stochastic Interpolants},
  author={Cuesta-Lazaro, Carolina and Bayer, Adrian E and Albergo, Michael S and others},
  booktitle={NeurIPS 2024 Workshop: Machine Learning and the Physical Sciences},
  year={2024}
}

@ARTICLE{dalal2023,
       author = {{Dalal}, Roohi and {Li}, Xiangchong and {Nicola}, Andrina and others},
        title = "{Hyper Suprime-Cam Year 3 results: Cosmology from cosmic shear power spectra}",
      journal = {\prd},
         year = 2023,
        month = dec,
       volume = {108},
       number = {12},
          eid = {123519},
        pages = {123519},
          doi = {10.1103/PhysRevD.108.123519},
archivePrefix = {arXiv},
       eprint = {2304.00701},
 primaryClass = {astro-ph.CO},
       adsurl = {https://ui.adsabs.harvard.edu/abs/2023PhRvD.108l3519D}
}

@ARTICLE{doux2022,
       author = {{Doux}, C. and {Jain}, B. and {Zeurcher}, D. and others},
        title = "{Dark energy survey year 3 results: cosmological constraints from the analysis of cosmic shear in harmonic space}",
      journal = {\mnras},
         year = 2022,
        month = sep,
       volume = {515},
       number = {2},
        pages = {1942-1972},
          doi = {10.1093/mnras/stac1826},
archivePrefix = {arXiv},
       eprint = {2203.07128},
 primaryClass = {astro-ph.CO},
       adsurl = {https://ui.adsabs.harvard.edu/abs/2022MNRAS.515.1942D}
}

@ARTICLE{fu2014,
        author = {{Fu}, Liping and {Kilbinger}, Martin and {Erben}, Thomas and
         {Heymans}, Catherine and others},
         title = "{CFHTLenS: cosmological constraints from a combination of cosmic shear two-point and three-point correlations}",
       journal = {\mnras},
          year = "2014",
         month = "Jul",
        volume = {441},
        number = {3},
         pages = {2725-2743},
           doi = {10.1093/mnras/stu754},
  primaryClass = {astro-ph.CO},
        adsurl = {https://ui.adsabs.harvard.edu/abs/2014MNRAS.441.2725F}
}

@ARTICLE{gatti2024a,
        author = {{Gatti}, M. and {Campailla}, G. and {Jeffrey}, N. and others},
         title = "{Dark Energy Survey Year 3 results: simulation-based cosmological inference with wavelet harmonics, scattering transforms, and moments of weak lensing mass maps II. Cosmological results}",
       journal = {arXiv e-prints},
          year = 2024,
         month = may,
           eid = {arXiv:2405.10881},
         pages = {arXiv:2405.10881},
           doi = {10.48550/arXiv.2405.10881},
archivePrefix = {arXiv},
        eprint = {2405.10881},
  primaryClass = {astro-ph.CO},
        adsurl = {https://ui.adsabs.harvard.edu/abs/2024arXiv240510881G}
}

@ARTICLE{gatti2024b,
        author = {{Gatti}, M. and {Jeffrey}, N. and {Whiteway}, L. and others},
         title = "{Dark Energy Survey Year 3 results: Simulation-based cosmological inference with wavelet harmonics, scattering transforms, and moments of weak lensing mass maps. Validation on simulations}",
       journal = {\prd},
          year = 2024,
         month = mar,
        volume = {109},
        number = {6},
           eid = {063534},
         pages = {063534},
           doi = {10.1103/PhysRevD.109.063534},
archivePrefix = {arXiv},
        eprint = {2310.17557},
  primaryClass = {astro-ph.CO},
        adsurl = {https://ui.adsabs.harvard.edu/abs/2024PhRvD.109f3534G}
}

@ARTICLE{gomes2025,
       author = {{Gomes}, R.~C.~H. and {Sugiyama}, S. and {Jain}, B. and others},
        title = "{Dark Energy Survey Year 3 Results: Cosmological constraints from second- and third-order shear statistics}",
      journal = {\prd},
         year = 2025,
        month = dec,
       volume = {112},
       number = {12},
          eid = {123515},
        pages = {123515},
          doi = {10.1103/sxlz-t9gb},
archivePrefix = {arXiv},
       eprint = {2508.14018},
 primaryClass = {astro-ph.CO},
       adsurl = {https://ui.adsabs.harvard.edu/abs/2025PhRvD.112l3515G}
}

@ARTICLE{grewal2022,
        author = {{Grewal}, Nisha and {Zuntz}, Joe and {Tr{\"o}ster}, Tilman and others},
         title = "{Minkowski Functionals in Joint Galaxy Clustering \& Weak Lensing Analyses}",
       journal = {The Open Journal of Astrophysics},
          year = 2022,
         month = aug,
        volume = {5},
        number = {1},
           eid = {13},
         pages = {13},
           doi = {10.21105/astro.2206.03877},
archivePrefix = {arXiv},
        eprint = {2206.03877},
  primaryClass = {astro-ph.CO},
        adsurl = {https://ui.adsabs.harvard.edu/abs/2022OJAp....5E..13G}
}

@ARTICLE{harnoisderaps2021,
       author = {{Harnois-D{\'e}raps}, Joachim and {Martinet}, Nicolas and {Castro}, Tiago and {Dolag}, Klaus and {Giblin}, Benjamin and {Heymans}, Catherine and {Hildebrandt}, Hendrik and {Xia}, Qianli},
        title = "{Cosmic shear cosmology beyond two-point statistics: a combined peak count and correlation function analysis of DES-Y1}",
      journal = {\mnras},
         year = 2021,
        month = sep,
       volume = {506},
       number = {2},
        pages = {1623-1650},
          doi = {10.1093/mnras/stab1623},
archivePrefix = {arXiv},
       eprint = {2012.02777},
 primaryClass = {astro-ph.CO},
       adsurl = {https://ui.adsabs.harvard.edu/abs/2021MNRAS.506.1623H}
}

@ARTICLE{hermans2021,
       author = {{Hermans}, Joeri and {Delaunoy}, Arnaud and {Rozet}, Fran{\c{c}}ois and others},
        title = "{A Trust Crisis In Simulation-Based Inference? Your Posterior Approximations Can Be Unfaithful}",
      journal = {arXiv e-prints},
         year = 2021,
        month = oct,
          eid = {arXiv:2110.06581},
        pages = {arXiv:2110.06581},
          doi = {10.48550/arXiv.2110.06581},
archivePrefix = {arXiv},
       eprint = {2110.06581},
 primaryClass = {stat.ML},
       adsurl = {https://ui.adsabs.harvard.edu/abs/2021arXiv211006581H}
}

@article{hoffman2014,
  author  = {Matthew D. Hoffman and Andrew Gelman},
  title   = {The No-U-Turn Sampler: Adaptively Setting Path Lengths in Hamiltonian Monte Carlo},
  journal = {Journal of Machine Learning Research},
  year    = {2014},
  volume  = {15},
  number  = {47},
  pages   = {1593--1623},
  url     = {http://jmlr.org/papers/v15/hoffman14a.html}
}

@article{jasche2010,
   title={Fast Hamiltonian sampling for large-scale structure inference: Fast Hamiltonian sampling},
   volume={407},
   ISSN={0035-8711},
   url={http://dx.doi.org/10.1111/j.1365-2966.2010.16897.x},
   DOI={10.1111/j.1365-2966.2010.16897.x},
   number={1},
   journal={Monthly Notices of the Royal Astronomical Society},
   publisher={Oxford University Press (OUP)},
   author={Jasche, Jens and Kitaura, Francisco S.},
   year={2010},
   month=jun, pages={29–42} }

@article{jasche2013,
   title={Bayesian physical reconstruction of initial conditions from large-scale structure surveys},
   volume={432},
   ISSN={1365-2966},
   url={http://dx.doi.org/10.1093/mnras/stt449},
   DOI={10.1093/mnras/stt449},
   number={2},
   journal={Monthly Notices of the Royal Astronomical Society},
   publisher={Oxford University Press (OUP)},
   author={Jasche, Jens and Wandelt, Benjamin D.},
   year={2013},
   month=apr, pages={894–913} }

@article{jeffrey2021,
  title={Likelihood-free inference with neural compression of DES SV weak lensing map statistics},
  author={Jeffrey, Niall and Alsing, Justin and Lanusse, Fran{\c{c}}ois},
  journal={Monthly Notices of the Royal Astronomical Society},
  volume={501},
  number={1},
  pages={954--969},
  year={2021},
  publisher={Oxford University Press}
}

@article{Kabalan2026, doi = {10.21105/joss.08852}, url = {https://doi.org/10.21105/joss.08852}, year = {2026}, publisher = {The Open Journal}, volume = {11}, number = {121}, pages = {8852}, author = {Kabalan, Wassim and Lanusse, François and Boucaud, Alexandre and Aubourg, Eric}, title = {jaxDecomp: JAX Library for 3D Domain Decomposition and Parallel FFTs}, journal = {Journal of Open Source Software} }

@article{kacprzak2016,
	Adsurl = {http://adsabs.harvard.edu/abs/2016MNRAS.463.3653K},
	Author = {{Kacprzak}, T. and {Kirk}, D. and {Friedrich}, O. and others},
	Collaboration = {DES Collaboration},
	Doi = {10.1093/mnras/stw2070},
	Journal = {\mnras},
	Month = dec,
	Pages = {3653-3673},
	Title = {{Cosmology constraints from shear peak statistics in Dark Energy Survey Science Verification data}},
	Volume = 463,
	Year = 2016}

@ARTICLE{kilbinger2015,
       author = {{Kilbinger}, Martin},
        title = "{Cosmology with cosmic shear observations: a review}",
      journal = {Reports on Progress in Physics},
         year = 2015,
        month = jul,
       volume = {78},
       number = {8},
          eid = {086901},
        pages = {086901},
          doi = {10.1088/0034-4885/78/8/086901},
archivePrefix = {arXiv},
       eprint = {1411.0115},
 primaryClass = {astro-ph.CO},
       adsurl = {https://ui.adsabs.harvard.edu/abs/2015RPPh...78h6901K}
}

@ARTICLE{kratochvil2012,
        author = {{Kratochvil}, Jan M. and {Lim}, Eugene A. and {Wang}, Sheng and
         {Haiman}, Zolt{\'a}n and others},
         title = "{Probing cosmology with weak lensing Minkowski functionals}",
       journal = {\prd},
          year = "2012",
         month = "May",
        volume = {85},
        number = {10},
           eid = {103513},
         pages = {103513},
           doi = {10.1103/PhysRevD.85.103513},
  primaryClass = {astro-ph.CO},
        adsurl = {https://ui.adsabs.harvard.edu/abs/2012PhRvD..85j3513K}
}

@ARTICLE{lanzieri2023,
       author = {{Lanzieri}, Denise and {Lanusse}, Fran{\c{c}}ois and {Modi}, Chirag and others},
        title = "{Forecasting the power of higher order weak-lensing statistics with automatically differentiable simulations}",
      journal = {\aap},
         year = 2023,
        month = nov,
       volume = {679},
          eid = {A61},
        pages = {A61},
          doi = {10.1051/0004-6361/202346888},
archivePrefix = {arXiv},
       eprint = {2305.07531},
 primaryClass = {astro-ph.IM},
       adsurl = {https://ui.adsabs.harvard.edu/abs/2023A&A...679A..61L}
}

@ARTICLE{lanzieri2025,
       author = {{Lanzieri}, Denise and {Zeghal}, Justine and {Lucas Makinen}, T. and others},
        title = "{Optimal neural summarization for full-field weak lensing cosmological implicit inference}",
      journal = {\aap},
         year = 2025,
        month = may,
       volume = {697},
          eid = {A162},
        pages = {A162},
          doi = {10.1051/0004-6361/202451535},
archivePrefix = {arXiv},
       eprint = {2407.10877},
 primaryClass = {astro-ph.CO},
       adsurl = {https://ui.adsabs.harvard.edu/abs/2025A&A...697A.162L}
}

@misc{lemos2023,
       title={Sampling-Based Accuracy Testing of Posterior Estimators for General Inference}, 
       author={Pablo Lemos and Adam Coogan and Yashar Hezaveh and others},
       year={2023},
       eprint={2302.03026},
       archivePrefix={arXiv},
       primaryClass={stat.ML},
       url={https://arxiv.org/abs/2302.03026}, 
}

@ARTICLE{list2024,
       author = {{List}, Florian and {Hahn}, Oliver},
        title = "{Perturbation-theory informed integrators for cosmological simulations}",
      journal = {Journal of Computational Physics},
         year = 2024,
        month = sep,
       volume = {513},
          eid = {113201},
        pages = {113201},
          doi = {10.1016/j.jcp.2024.113201},
archivePrefix = {arXiv},
       eprint = {2301.09655},
 primaryClass = {astro-ph.CO},
       adsurl = {https://ui.adsabs.harvard.edu/abs/2024JCoPh.51313201L}
}

@ARTICLE{list2025,
       author = {{List}, Florian and {Hahn}, Oliver and {Fl{\"o}ss}, Thomas and others},
        title = "{DISCO-DJ II: a differentiable particle-mesh code for cosmology}",
      journal = {arXiv e-prints},
         year = 2025,
        month = oct,
          eid = {arXiv:2510.05206},
        pages = {arXiv:2510.05206},
          doi = {10.48550/arXiv.2510.05206},
archivePrefix = {arXiv},
       eprint = {2510.05206},
 primaryClass = {astro-ph.CO},
       adsurl = {https://ui.adsabs.harvard.edu/abs/2025arXiv251005206L}
}

@ARTICLE{liux2015,
       author = {{Liu}, Xiangkun and {Pan}, Chuzhong and {Li}, Ran and others},
        title = "{Cosmological constraints from weak lensing peak statistics with Canada-France-Hawaii Telescope Stripe 82 Survey}",
      journal = {\mnras},
         year = 2015,
        month = jul,
       volume = {450},
       number = {3},
        pages = {2888-2902},
          doi = {10.1093/mnras/stv784},
archivePrefix = {arXiv},
       eprint = {1412.3683},
 primaryClass = {astro-ph.CO},
       adsurl = {https://ui.adsabs.harvard.edu/abs/2015MNRAS.450.2888L}
}

@ARTICLE{mandelbaum2018,
       author = {{Mandelbaum}, Rachel},
        title = "{Weak Lensing for Precision Cosmology}",
      journal = {\araa},
         year = 2018,
        month = sep,
       volume = {56},
        pages = {393-433},
          doi = {10.1146/annurev-astro-081817-051928},
archivePrefix = {arXiv},
       eprint = {1710.03235},
 primaryClass = {astro-ph.CO},
       adsurl = {https://ui.adsabs.harvard.edu/abs/2018ARA&A..56..393M}
}

@ARTICLE{marques2024,
       author = {{Marques}, Gabriela A. and {Liu}, Jia and {Shirasaki}, Masato and others},
        title = "{Cosmology from weak lensing peaks and minima with Subaru Hyper Suprime-Cam Survey first-year data}",
      journal = {\mnras},
         year = 2024,
        month = mar,
       volume = {528},
       number = {3},
        pages = {4513-4527},
          doi = {10.1093/mnras/stae098},
archivePrefix = {arXiv},
       eprint = {2308.10866},
 primaryClass = {astro-ph.CO},
       adsurl = {https://ui.adsabs.harvard.edu/abs/2024MNRAS.528.4513M}
}

@INCOLLECTION{neal2011,
        author = {{Neal}, Radford},
         title = "{MCMC Using Hamiltonian Dynamics}",
     booktitle = {Handbook of Markov Chain Monte Carlo},
          year = 2011,
         pages = {113-162},
           doi = {10.1201/b10905},
        adsurl = {https://ui.adsabs.harvard.edu/abs/2011hmcm.book..113N}
}

@ARTICLE{novaes2025,
       author = {{Novaes}, Camila P. and {Thiele}, Leander and {Armijo}, Joaquin and others},
        title = "{Cosmology from HSC Y1 weak lensing data with combined higher-order statistics and simulation-based inference}",
      journal = {\prd},
         year = 2025,
        month = apr,
       volume = {111},
       number = {8},
          eid = {083510},
        pages = {083510},
          doi = {10.1103/PhysRevD.111.083510},
archivePrefix = {arXiv},
       eprint = {2409.01301},
 primaryClass = {astro-ph.CO},
       adsurl = {https://ui.adsabs.harvard.edu/abs/2025PhRvD.111h3510N}
}

@article{papamakarios2016,
  title={Fast $\varepsilon$-free inference of simulation models with bayesian conditional density estimation},
  author={Papamakarios, George and Murray, Iain},
  journal={Advances in neural information processing systems},
  volume={29},
  year={2016}
}

@InProceedings{papamakarios2019,
  title = 	 {Sequential Neural Likelihood: Fast Likelihood-free Inference with Autoregressive Flows},
  author =       {Papamakarios, George and Sterratt, David and Murray, Iain},
  booktitle = 	 {Proceedings of the Twenty-Second International Conference on Artificial Intelligence and Statistics},
  pages = 	 {837--848},
  year = 	 {2019},
  editor = 	 {Chaudhuri, Kamalika and Sugiyama, Masashi},
  volume = 	 {89},
  series = 	 {Proceedings of Machine Learning Research},
  month = 	 {16--18 Apr},
  publisher =    {PMLR},
  url = 	 {https://proceedings.mlr.press/v89/papamakarios19a.html}
}

@ARTICLE{porqueres2022,
        author = {{Porqueres}, Natalia and {Heavens}, Alan and {Mortlock}, Daniel and others},
         title = "{Lifting weak lensing degeneracies with a field-based likelihood}",
       journal = {\mnras},
          year = 2022,
         month = jan,
        volume = {509},
        number = {3},
         pages = {3194-3202},
           doi = {10.1093/mnras/stab3234},
archivePrefix = {arXiv},
        eprint = {2108.04825},
  primaryClass = {astro-ph.CO},
        adsurl = {https://ui.adsabs.harvard.edu/abs/2022MNRAS.509.3194P}
}

@ARTICLE{porqueres2023,
        author = {{Porqueres}, Natalia and {Heavens}, Alan and {Mortlock}, Daniel and others},
         title = "{Field-level inference of cosmic shear with intrinsic alignments and baryons}",
       journal = {arXiv e-prints},
          year = 2023,
         month = apr,
           eid = {arXiv:2304.04785},
         pages = {arXiv:2304.04785},
           doi = {10.48550/arXiv.2304.04785},
archivePrefix = {arXiv},
        eprint = {2304.04785},
  primaryClass = {astro-ph.CO},
        adsurl = {https://ui.adsabs.harvard.edu/abs/2023arXiv230404785P}
}

@ARTICLE{rampf2025,
       author = {{Rampf}, Cornelius and {List}, Florian and {Hahn}, Oliver},
        title = "{BULLFROG: multi-step perturbation theory as a time integrator for cosmological simulations}",
      journal = {\jcap},
         year = 2025,
        month = feb,
       volume = {2025},
       number = {2},
          eid = {020},
        pages = {020},
          doi = {10.1088/1475-7516/2025/02/020},
archivePrefix = {arXiv},
       eprint = {2409.19049},
 primaryClass = {astro-ph.CO},
       adsurl = {https://ui.adsabs.harvard.edu/abs/2025JCAP...02..020R}
}

@ARTICLE{regaldosaintblancard2023,
        author = {{R{\'e}galdo-Saint Blancard}, Bruno and {Allys}, Erwan and {Auclair}, Constant and others},
         title = "{Generative Models of Multichannel Data from a Single Example-Application to Dust Emission}",
       journal = {\apj},
          year = 2023,
         month = jan,
        volume = {943},
        number = {1},
           eid = {9},
         pages = {9},
           doi = {10.3847/1538-4357/aca538},
archivePrefix = {arXiv},
        eprint = {2208.03538},
  primaryClass = {astro-ph.CO},
        adsurl = {https://ui.adsabs.harvard.edu/abs/2023ApJ...943....9R}
}

@ARTICLE{secco2022,
        author = {{Secco}, L.~F. and {Jarvis}, M. and {Jain}, B. and others},
         title = "{Dark Energy Survey Year 3 Results: Three-point shear correlations and mass aperture moments}",
       journal = {\prd},
          year = 2022,
         month = may,
        volume = {105},
        number = {10},
           eid = {103537},
         pages = {103537},
           doi = {10.1103/PhysRevD.105.103537},
archivePrefix = {arXiv},
        eprint = {2201.05227},
  primaryClass = {astro-ph.CO},
        adsurl = {https://ui.adsabs.harvard.edu/abs/2022PhRvD.105j3537S}
}

@ARTICLE{semboloni2011,
        author = {{Semboloni}, Elisabetta and {Schrabback}, Tim and
         {van Waerbeke}, Ludovic and {Vafaei}, Sanaz and others},
         title = "{Weak lensing from space: first cosmological constraints from three-point shear statistics}",
       journal = {\mnras},
          year = "2011",
         month = "Jan",
        volume = {410},
        number = {1},
         pages = {143-160},
           doi = {10.1111/j.1365-2966.2010.17430.x},
  primaryClass = {astro-ph.CO},
        adsurl = {https://ui.adsabs.harvard.edu/abs/2011MNRAS.410..143S}
}

@ARTICLE{sugiyama2025,
      title={Cosmology from a joint analysis of second and third order shear statistics with Subaru Hyper Suprime-Cam Year 3 data}, 
      author={Sunao Sugiyama and Rafael C. H. Gomes and Bhuvnesh Jain},
      year={2025},
      eprint={2508.14019},
      archivePrefix={arXiv},
      primaryClass={astro-ph.CO},
      url={https://arxiv.org/abs/2508.14019}, 
}

@article{thomas2022likelihood,
  title={Likelihood-free inference by ratio estimation},
  author={Thomas, Owen and Dutta, Ritabrata and Corander, Jukka and others},
  journal={Bayesian Analysis},
  volume={17},
  number={1},
  pages={1--31},
  year={2022},
  publisher={International Society for Bayesian Analysis}
}

@ARTICLE{vonwietersheimkramsta2024,
        author = {{von Wietersheim-Kramsta}, Maximilian and {Lin}, Kiyam and {Tessore}, Nicolas and others},
         title = "{KiDS-SBI: Simulation-Based Inference Analysis of KiDS-1000 Cosmic Shear}",
       journal = {arXiv e-prints},
          year = 2024,
         month = apr,
           eid = {arXiv:2404.15402},
         pages = {arXiv:2404.15402},
           doi = {10.48550/arXiv.2404.15402},
archivePrefix = {arXiv},
        eprint = {2404.15402},
  primaryClass = {astro-ph.CO},
        adsurl = {https://ui.adsabs.harvard.edu/abs/2024arXiv240415402V}
}

@article{wood2010statistical,
  title={Statistical inference for noisy nonlinear ecological dynamic systems},
  author={Wood, Simon N},
  journal={Nature},
  volume={466},
  number={7310},
  pages={1102--1104},
  year={2010},
  publisher={Nature Publishing Group UK London}
}

@ARTICLE{wright2025,
       author = {{Wright}, Angus H. and {St{\"o}lzner}, Benjamin and {Asgari}, Marika and others},
        title = "{KiDS-Legacy: Cosmological constraints from cosmic shear with the complete Kilo-Degree Survey}",
      journal = {\aap},
         year = 2025,
        month = nov,
       volume = {703},
          eid = {A158},
        pages = {A158},
          doi = {10.1051/0004-6361/202554908},
archivePrefix = {arXiv},
       eprint = {2503.19441},
 primaryClass = {astro-ph.CO},
       adsurl = {https://ui.adsabs.harvard.edu/abs/2025A&A...703A.158W}
}

@ARTICLE{zeghal2025,
        author = {{Zeghal}, Justine and {Lanzieri}, Denise and {Lanusse}, Fran{\c{c}}ois and others},
         title = "{Simulation-based inference benchmark for weak lensing cosmology}",
       journal = {\aap},
          year = 2025,
         month = jul,
        volume = {699},
           eid = {A327},
         pages = {A327},
           doi = {10.1051/0004-6361/202452410},
archivePrefix = {arXiv},
        eprint = {2409.17975},
  primaryClass = {astro-ph.CO},
        adsurl = {https://ui.adsabs.harvard.edu/abs/2025A&A...699A.327Z}
}

@ARTICLE{zehgal2025b,
       author = {{Zeghal}, Justine and {Remy}, Benjamin and {Hezaveh}, Yashar and others},
        title = "{Bridging Simulators with Conditional Optimal Transport}",
      journal = {arXiv e-prints},
         year = 2025,
        month = oct,
          eid = {arXiv:2510.24631},
        pages = {arXiv:2510.24631},
          doi = {10.48550/arXiv.2510.24631},
archivePrefix = {arXiv},
       eprint = {2510.24631},
 primaryClass = {astro-ph.CO},
       adsurl = {https://ui.adsabs.harvard.edu/abs/2025arXiv251024631Z}
}

\clearpage
\onecolumngrid
\clearpage

\appendix

\section{TARP}\label{app:tarp}
\begin{figure*}
\centering
\includegraphics[width=0.99\linewidth]{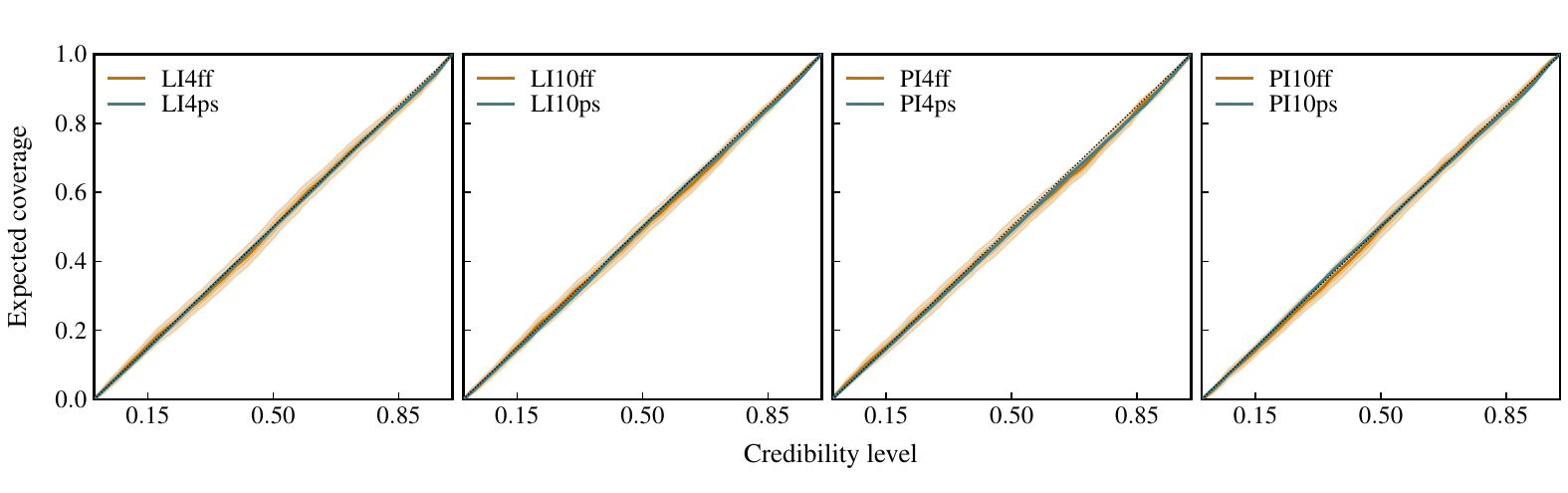}
\caption{TARP calibration for posterior uncertainty. Each panel shows the empirical expected coverage as a function of nominal credibility level for the configuration labeled in the legend. The dashed diagonal line indicates perfect calibration (expected coverage equals credibility); deviations below (above) the line indicate overconfident (underconfident) posteriors. Solid curves show the mean over simulations, with shaded bands indicating the scatter across realizations (shown only for full-field variants in orange).\\}
\label{fig:tarp}
\end{figure*}

We validate the coverage of our implicit-inference posteriors using TARP. We draw parameter values $\vec{\theta}_{i}$ from the prior, generate a mock map $x_i$ for each, and run our inference pipeline to obtain the estimated posterior $\hat{p}(\vec{\theta}\,|\,x_i)$. At each nominal level $\alpha$, we then measure the fraction of true values $\vec{\theta}_{i}$ lying within the corresponding $\alpha$-credible region. This fraction is a Monte Carlo estimate of the expected coverage probability,
\begin{equation}
{\rm ECP}(\alpha)
=
\mathbb{E}_{p(\theta, x)}
\left[
\mathbf{1}\!\left\{\vec{\theta} \in \mathcal{G}(\hat{p}, \alpha, x)\right\}
\right],
\end{equation}
where $\theta$ is the true value used to generate $x$, and
$\mathcal{G}(\hat{p}, \alpha, x)$ is the $\alpha$-credible region of the
posterior. A posterior is well calibrated if ${\rm ECP}(\alpha) = \alpha$ for all $\alpha$ (e.g., the true parameters lie within the $68\%$ credible region in $68\%$ of the simulations), or equivalently, the coverage curve should follow the identity line. This test requires inferring posteriors for many simulated maps, which is impractical for explicit inference, since each posterior requires a full sampling run, but straightforward for implicit inference, where posterior evaluation is fast once the density estimator is trained. As shown in Figure~\ref{fig:tarp}, the coverage curves for both the power-spectrum and the full-field implicit analyses closely follow the identity line at all credibility levels, indicating that the posteriors are well calibrated.\\

\section{Sampling Stiff Cosmology--Latent field Directions}\label{app:precond}
The dominant stiffness in the field-level posterior arises from the strong posterior correlation between $\vec{\theta}$ and $\vec{u}$, induced by the likelihood. The likelihood constrains the forward-model prediction $\mathcal{M}(\vec{\theta}, \vec{u})$ to remain close to the fixed observed field $\vec{d}$: any shift in $\vec{\theta}$ (which modifies both the mapping from $\vec{u}$ to the physical initial density field and the subsequent forward evolution, including the growth history, lensing geometry, and projected convergence field) must therefore be compensated by a coherent adjustment in $\vec{u}$ to keep the output map consistent with $\vec{d}$. As a result, the high-probability region of the posterior forms a narrow ridge in the joint $(\vec{\theta},\vec{u})$ parameter space. Moving in $\vec{\theta}$ while holding $\vec{u}$ fixed quickly leaves this region, making cosmological exploration inefficient unless the sampler can follow the coupled directions in which changes in $\vec{\theta}$ are compensated by corresponding changes in $\vec{u}$.

Although $\vec{u}$ is high-dimensional, most modes are only weakly constrained by the data. Weak lensing convergence is a projected quantity, obtained by integrating the matter field along the line of sight with a broad lensing kernel. This projection averages over small-scale fluctuations, especially those that vary rapidly in the radial direction, so many modes of $\vec{u}$ have only a weak effect on the observed convergence map. Their posterior distribution is therefore dominated by the Gaussian prior. The coupling to $\vec{\theta}$ is consequently concentrated in a much smaller subspace of $\vec{u}$. This subspace is not generally spanned by individual Fourier modes, but rather by combinations of modes that collectively contribute to the lensing map. These combinations can partially compensate for changes in $\Omega_{\rm m}$ or $S_{8}^{\rm lin}$, allowing the model to remain close to the observed field even as those parameters are varied.

This low-dimensional coupling structure is useful for HMC/NUTS sampling. Efficient sampling benefits from proposals that are aligned with the coupled directions of the posterior, rather than steps that cut across them. In HMC, this information is encoded in the inverse mass matrix: the diagonal elements set the relative scale of different parameter directions, while the off-diagonal elements encode correlations between directions and allow the sampler to propose coherent joint moves. A dense inverse mass matrix over the full $\mathcal{O}(10^7)$-dimensional parameter space would be computationally intractable. However, the argument above suggests that the important off-diagonal structure is concentrated in the much smaller joint space spanned by $\vec{\theta}$ and the cosmology-coupled directions in $\vec{u}$. We therefore use a dense inverse mass matrix block only in this reduced space, while treating the remaining weakly coupled modes with a diagonal inverse mass matrix. This captures the dominant posterior degeneracies while keeping the sampler computationally tractable.

To construct the inverse mass matrix, we first identify which coherent directions of $\vec{u}$ respond most strongly to changes in cosmology. We do this by probing how $\vec{u}$ changes as the cosmological parameters are moved across the prior. We begin by selecting $8\times8=64$ cosmological parameter points inside the prior plane. At each point $\vec{\theta}_j$, the cosmology is held fixed and $\vec{u}$ is relaxed toward the conditional MAP,
\begin{equation}\label{eq:conditional_ic_map}
\hat{u}_j=\arg\min_u U(\vec{\theta}_j,\vec{u}),
\end{equation}
where $U(\vec{\theta},\vec{u})$ is the posterior potential, i.e., the negative log posterior conditioned on the data map used for inference. Since $\vec{\theta}_j$ is fixed during this solve, the optimization is only over $\vec{u}$. The conditional MAP $\hat{u}_j$ therefore describes the preferred adjustment of the whitened latent field at cosmology $\vec{\theta}_j$, chosen to keep the forward-modeled convergence map close to the data convergence map  while remaining consistent with the latent field prior. At each conditional MAP, the latent field gradient is approximately zero,
\begin{equation}
\nabla_u U(\vec{\theta}_j,\hat{u}_j) \simeq 0,
\end{equation}
because the initial conditions have been optimized at fixed cosmology.  We then perturb one cosmological parameter at a time while holding the optimized latent field fixed. For each cosmological parameter $\theta_\alpha$, perturbed by $\delta{\theta}_{\alpha}$, we define this latent-field gradient response as $x_{j\alpha}$:
\begin{equation}\label{eq:coupling_direction}
\nabla_u U(\vec{\theta}_{j} + \delta\theta_\alpha, \hat{u}_{j})\approx\delta\theta_\alpha x_{j\alpha},
\qquad
x_{j\alpha} =\frac{\partial}{\partial \theta_\alpha}\nabla_u U(\vec{\theta}, \vec{u})
\bigg|_{\vec{\theta}=\vec{\theta}_{j},\, \vec{u}=\hat{u}_{j}}.
\end{equation}
Here $x_{j\alpha}$ is a vector with the same dimensionality as $\vec{u}$. Its components describe how the posterior force on each latent-field mode changes when the cosmological parameter $\theta_\alpha$ is perturbed by $\delta\theta_\alpha$. Thus, $x_{j\alpha}$ identifies a latent-field direction that is coupled to the cosmological parameter $\theta_\alpha$.

We repeat this procedure at all 64 grid points for each parameter, $\theta_{\alpha} \in \{\Omega_{\rm m}, S_8^{\rm lin}\}$, giving 128 response vectors $\vec{x}_{j\alpha}$. We normalize each response vector and stack them as the rows of a matrix $X_{\rm coup}\in\mathbb{R}^{128\times \dim(\vec{u})}$. The singular value decomposition
\begin{equation}
X_{\rm coup} = W \Sigma V^{T}
\end{equation}
provides an orthonormal basis for the subspace spanned by these response vectors. We take the leading $k$ right singular vectors, $V_k = [v_1 \cdots v_k]$, which capture the largest common components of the responses, to define the directions in the latent-field coordinates that are most strongly coupled to the cosmological parameters. We complete this basis with an orthonormal complement $V_\perp$ to form an orthogonal rotation $Q = [\,V_k\; V_\perp\,]$ and define rotated latent coordinates
\begin{equation}
\begin{bmatrix}
\vec{a}\\
\vec{b}
\end{bmatrix}
= Q^{T}\vec{u},
\label{eq:householder_rotation}
\end{equation}
where $\vec{a}$ contains the coordinates along the cosmology-coupled directions and $\vec{b}$ contains the remaining latent-field coordinates. We then sample in $(\vec{a},\vec{b})$ rather than in $\vec{u}$. In these coordinates, we use a block-structured inverse mass matrix, following the convention used by \texttt{NumPyro}. The block containing the cosmological parameters $\vec{\theta}$ and the coupled coordinates $\vec{a}$ is treated densely, while the remaining latent-field coordinates $\vec{b}$ are assigned an analytic Fourier-space diagonal inverse mass matrix. Writing $\Lambda \equiv M^{-1}$, the inverse mass matrix used by NUTS has the form
\begin{equation}
\Lambda =
\begin{pmatrix}
\begin{bmatrix}
\Lambda_{\theta\theta} & \Lambda_{\theta a} \\
\Lambda_{a\theta} & \Lambda_{aa}
\end{bmatrix}
& 0 \\
0 & \Lambda_b
\end{pmatrix}.
\end{equation}
The upper-left submatrix is the dense block for the joint coordinates $(\vec{\theta},\vec{a})$, with dimension $(N_\theta+k)\times(N_\theta+k)$, where $N_\theta=2$ and $k$ is the number of retained coupling directions. We use $k=12$ for LPT and $k=16$ for PM. The remaining block, $\Lambda_b$, contains the rest of the latent field and is kept diagonal, with entries given by the analytic Fourier-space inverse mass. After fixing this block structure, we run a short 150-step warmup chain in the rotated coordinates. During this warmup, NUTS adapts only the small dense block, while $\Lambda_b$ is held fixed. This confines mass matrix adaptation to the low-dimensional subspace where cosmology and latent-field parameters are most strongly coupled, while the high-dimensional complement is held fixed.

Using the inverse mass matrix obtained from the warmup described above, NUTS naturally reaches a U-turn at tree depths of $7$ and $8$ for the LPT and PM forward models, respectively, while we set the maximum tree depth to $9$. Although this procedure does not guarantee an optimal inverse mass matrix, it provides a practical way to construct one that captures enough of the dominant couplings between the cosmological parameters and the latent field to allow NUTS to move along the posterior ridge and explore the relevant parameter space. In contrast, when sampling directly in the original coordinates without the rotation and block-structured inverse mass matrix, NUTS frequently reaches the maximum tree depth before identifying a U-turn. In this unrotated case, the trajectories are truncated and explore only a small region of parameter space, making full posterior exploration computationally impractical.

\section{Alternate compression: CNN}\label{app:cnn}
\begin{figure}
\centering
\includegraphics[width=0.98\linewidth]{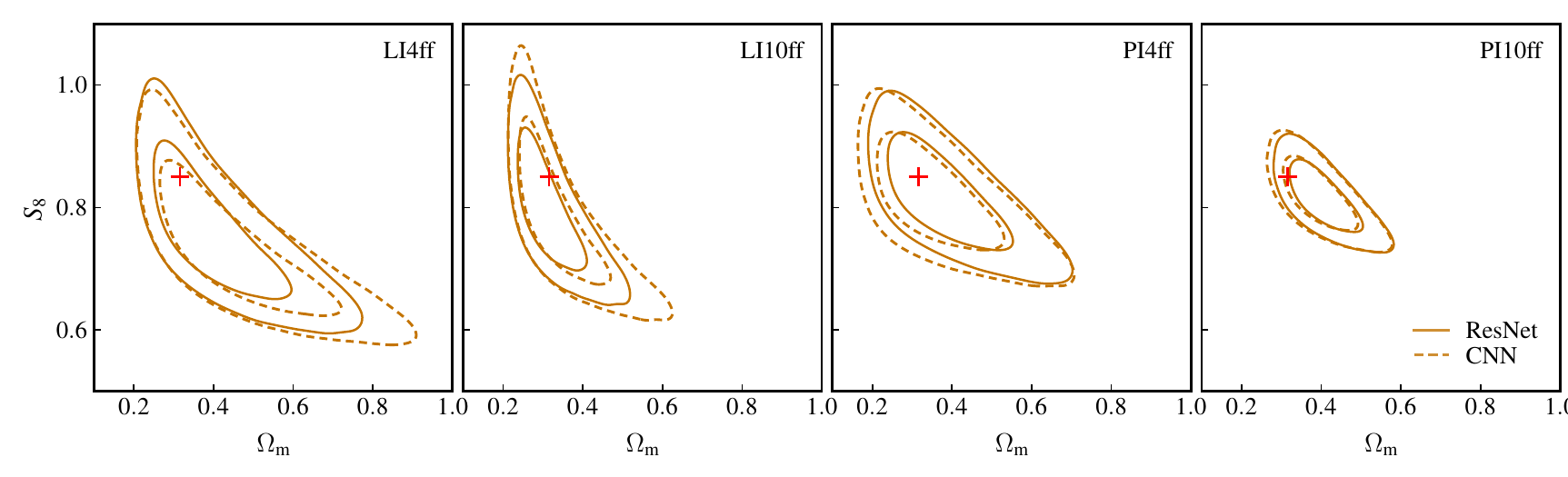}
\caption{Comparison of implicit field-level constraints obtained using the baseline ResNet compressor and a simpler CNN compressor. Both architectures recover similar degeneracy directions, but the ResNet compressor gives consistently tighter constraints.\\}
\label{fig:resnet_vs_cnn}
\end{figure}

In the implicit approach, the neural compressor determines what information is passed from the map to the posterior estimator. As a cross-check on this compression step, we compare our fiducial ResNet-based summaries with those from a simpler CNN architecture consisting of three convolutional layers with 32, 64, and 128 filters, respectively, each using a $3\times3$ kernel and stride 2, followed by average pooling and a dense layer mapping to the 32-dimensional summary. We show the resulting posterior comparison in Figure~\ref{fig:resnet_vs_cnn}. The two architectures recover broadly similar degeneracy directions, giving confidence that the ResNet compressor is capturing the dominant cosmological information in the maps. However, the CNN contours are generally broader, showing that the architecture and training procedure affect the quality of the learned summaries, even when the training objective can in principle learn sufficient statistics.

\end{document}